\documentclass[12pt,preprint]{aastex}

\newcommand{\Lya}{\hbox{Ly$\alpha$}}

\newcommand{\Jup}{\hbox{$J^\prime$}}
\newcommand{\Jlo}{\hbox{$J^{\prime\prime}$}}
\newcommand{\Vup}{\hbox{$v^\prime$}}
\newcommand{\Vlo}{\hbox{$v^{\prime\prime}$}}
\newcommand{\Elo}{\hbox{$E^{\prime\prime}$}}
\newcommand{\Ehi}{\hbox{$E^{\prime}$}}
\newcommand{\kms}{\hbox{km s$^{-1}$}}
\newcommand{\IUE}{\textit{IUE}}
\newcommand{\HST}{\textit{HST}}
\newcommand{\STIS}{STIS}            
\newcommand{\FUSE}{\textit{FUSE}}
\newcommand{\IDL}{IDL}              

\slugcomment{Accepted to ApJ}
\shorttitle{H$_2$ Fluorescence in the UV Spectrum of TW Hya}
\shortauthors{Herczeg et al.}

\begin{document}

\title{The FUV spectrum of TW Hya. I. Observations of H$_2$ Fluorescence$^1$}

\author{Gregory J. Herczeg and Jeffrey L. Linsky}
\affil{JILA, University of Colorado and NIST, Boulder, CO 80309--0440}
\email{gregoryh@casa.colorado.edu,jlinsky@jila.colorado.edu}

\author{Jeff A. Valenti}
\affil{Space Telescope Science Institute, Baltimore, MD 21218}
\email{valenti@stsci.edu}

\author{Christopher M. Johns-Krull}
\affil{Department of Physics and Astronomy, Rice University, 6100 Main
       St. MS-108, Houston, TX 77005}
\email{cmj@rice.edu}

\and

\author{Brian E. Wood}
\affil{JILA, University of Colorado and NIST, Boulder, CO 80309--0440}
\email{woodb@marmoset.colorado.edu}

\footnotetext[1]{Based on observations with the NASA/ESA Hubble Space
Telescope, obtained at the Space Telescope Science Institute, which is
operated by the Association of Universities for Research in Astronomy,
Inc. under NASA contract NAS5-26555.}

\begin{abstract}
We observed the classical T Tauri star TW Hya with \textit{HST}/STIS
using the E140M grating, from 1150--1700 \AA,
with the E230M grating, from 2200--2900 \AA, and with \FUSE\ from
900--1180 \AA.  Emission in 143 Lyman-band H$_2$ lines
representing 19 progressions dominates the spectral region from
1250--1650 \AA.
The total H$_2$ emission line flux is $1.94 \times 10^{-12}$ erg cm$^{-2}$
s$^{-1}$, which corresponds to $1.90\times10^{-4}$ $L_\odot$ at TW Hya's
distance of 56 pc.  A
broad stellar \Lya\ line photoexcites the H$_2$ from excited
rovibrational levels of the ground electronic state to 
excited electronic states.
The \ion{C}{2} 1335 \AA\ doublet, \ion{C}{3} 1175 \AA\ multiplet, and
\ion{C}{4} 1550 \AA\ doublet also electronically excite H$_2$.  
The
velocity shift of the H$_2$ lines is consistent with the photospheric
radial velocity of TW Hya, and the emission is not spatially extended
beyond the $0\farcs05$ resolution of \textit{HST}.  
The H$_2$ lines have an intrinsic FWHM of $11.91\pm0.16$ \kms.
One H$_2$ line is
significantly weaker than predicted by this model because of \ion{C}{2}
wind absorption.  We also do not observe any H$_2$ absorption against the
stellar \Lya\ profile.  From these results, we conclude that the H$_2$ 
emission is
more consistent with an origin in a disk rather than in an outflow or
circumstellar shell.
We also analyze the hot accretion-region lines 
(e.g., \ion{C}{4}, \ion{Si}{4},
\ion{O}{6}) of TW Hya, which
are formed at the accretion shock, and discuss some reasons why Si lines
 appear
significantly weaker than other TR region lines.
\end{abstract}


%
%

\section{INTRODUCTION}

T Tauri stars (TTSs) are young ($< 10$ Myr), roughly solar mass stars
that have only recently emerged from their natal molecular clouds to
become optically visible \citep[see reviews by][]{Ber89,Fei99}.  The
presence of disks around classical TTSs (CTTSs) is now well established
observationally.  Millimeter wave measurements of the dust continuum
from the outer disk can be used to estimate the disk mass, assuming a
dust to gas ratio (see review by Zuckerman 2001).  Such estimates typically yield masses of $\sim
0.02$ M$_\odot$, consistent with the minimum mass for the solar nebula
\citep[e.g.,][]{Bec90}.  Spectral line observations can be used to
study the inner regions of the disk through their kinematic
signatures.  This has been done successfully for very young embedded
sources using the infrared (IR) bands of CO \citep*{Car01,Thi01,Naj96,Cha95},
but the results are confusing for the CTTSs that show CO band emission
\citep{Thi01,Cha95}.

Molecular hydrogen is expected to be $\sim 10^4$ times more
abundant than other gas tracers such as CO in the disks surrounding
young stars.  Depending on the density,
H$_2$ can survive at temperatures of up to 4000 K \citep[e.g.,][]{Wil00a},
and can self-shield well against UV radiation fields, making it an
excellent diagnostic of disks around young stars.  Good probes of gas
in circumstellar disks are needed to determine the lifetime of gas in
these environments where planets likely form.  Most studies of disks to
date have relied on the IR to millimeter wavelength spectral energy
distribution (SED) of these young stars.  These studies are mainly
sensitive to micron-sized dust and have shown that the typical survival
time for this dust is only a few million years
\citep*{Str93,Wol96,Hil99,Alv00}.  This does not mean, however, that the
disks around these stars disappear on this short timescale.  Theories
of giant planet formation suggest that after the dust collects into
larger particles (and no longer shows up in studies of the SED), the
gas is still present in the disk for some time before it accretes onto
the planets \citep*[e.g.,][]{Wuc00}.  Indeed, if for most CTTSs the gas
disappears from the disk on the same timescale as the dust does, it may
be quite difficult to form giant gas planets.

Molecular hydrogen around a CTTS was first detected by 2.12183 $\mu$m
1-0 S(1) line emission in the spectrum of T Tau \citep{Bec78}.
\citet{Bro81} discovered ultraviolet emission lines of H$_2$ in an
\IUE\ spectrum of T Tau and identified them as fluorescent lines pumped
by \Lya, which were previously seen in sunspot spectra.  In both cases, it
was apparent that the molecular hydrogen emission from T Tau is
spatially extended.  Subsequently, H$_2$ has been studied often around
CTTSs in both the IR and UV. \citet{Car90} surveyed young stellar
objects in the IR, finding H$_2$ emission in 4 CTTSs.  The location of
this H$_2$ emission is very important.  Recent ground-based images of T
Tau show that the IR emission is quite extended, reaching to
20\arcsec\ from the star \citetext{\citealp{Van94}; see also
\citealp*{Her96,Her97}}.  The extended IR H$_2$ emission seen around T
Tau and other young embedded sources like L1448 \citep*{Bal93} is
interpreted as shock heated emission from the interaction of the
stellar/disk wind/jet with the surrounding cloud material
\citep[see review by][]{Rei01}.
\citet{Thi01} used ISO to detect H$_2$ emission in the IR from 22 CTTSs,
Herbig Ae/Be stars, and debris-disk stars,
 but with a large aperture extending over
20\arcsec\ this emission may include both disk emission and spatially
extended H$_2$ emission from outflows.  They assumed based on
the strength of on-source versus off-source CO emission that 
the bulk of the emission occurred in the circumstellar disks.  However,
\citet{Ric01} attemped to detect on-source H$_2$ emission from five of
these sources using a small slit, but failed to detect any significant
H$_2$ emission.

Magnetic quadrupole transitions of H$_2$ in the IR are very weak, and
electric dipole transitions within the ground electronic state 
are forbidden because H$_2$ has no net
dipole moment.  In the UV, the large oscillator strengths of the electronic
Lyman-band (B-X) transitions lead to strong emission and absorption
lines via transitions to and from the excited electronic state.
\citet*{Val00} observed H$_2$ emission in the UV around 13 of 32 TTSs
studied with \IUE.  \citet{Val00}  go on to argue that the limited sensitivity of \IUE\ may be
the only reason that H$_2$ was not detected in the remaining 19 TTS.  \citet*{Joh00} attributed UV pumping to the
excitation of H$_2$, because the observed flux in H$_2$ lines scales
with the total UV flux.  
The UV H$_2$ lines
of T Tau are spatially extended beyond the disks thought to
be in this system and may have substantial contributions from
outflows.  However, as described below, we do not think that the H$_2$
lines from TW Hya arise in a molecular outflow.

More recently, \citet{Ard01} detected fluorescent H$_2$ around a sample
of CTTSs with \HST/GHRS and attributed the pumping to the red wing of \Lya.  \citet{Rob01}
and \citet{Her01} presented observations of circumstellar H$_2$
absorption against the \ion{O}{6} profiles of the Herbig Ae/Be stars AB
Aur and DX Cha, as observed with the Far Ultraviolet Spectrograph
Explorer (\FUSE).  \citet{Her01} identified H$_2$ emission in
\FUSE\ observations of TW Hya and V4046 Sgr in two Lyman-band H$_2$
lines, but did not detect 
significant H$_2$
absorption against \ion{O}{6} for these two stars.

TW Hya is the most prominent member of the recently named TW Hydrae
association \citep[TWA --][]{Web99} located at a distance of $56\pm8$ pc
\citep{Wic98}, making it the closest known CTTS to the solar
system. \citet{Web99} estimated that TW Hya is a $0.7$ 
$M_\odot$ star with an age of 10 Myr based on evolutionary tracks of
\citet{Dan97}.  Most other members of the TWA are naked T Tauri stars
(NTTSs - TTSs which lack close circumstellar disks), which confirms
that TW Hya is quite old for a CTTS.
TW Hya shows strong H$\alpha$ emission \citep{Ruc83,Web99} produced in
the accretion column,
and an IR
excess \citep{Jay99} produced by dust emission from a disk.  
The mass accretion rate has been measured to be
from $5-100\times10^{-10}M_\odot$ yr$^{-1}$ \citep{Muz00,Ale01}.
By measuring the width of magnetically sensitive
lines, \citet{Joh01a} estimated that the mean magnetic field of TW Hya is
of order 3 kG.
Recent imaging of TW Hya shows that its disk is viewed very nearly
face-on \citep{Wei99,Kri00}.  D. Potter \citetext{2001, private
communication} determined the inclination angle $i=10\pm5^\circ$
using IR polarimetry,
while \citet{Ale01} estimated an inclination of $i=17.5\pm4.5^\circ$ 
by analyzing time-series profiles of H$\alpha$.
The outer radius of the disk extends
more than 225 AU \citep{Kri00,Tri01,Wei01} from the star, and its inner
truncation radius is about 0.3 AU \citep{Tri01}.  \citet{Dal01} found
evidence from comparing the observed spectral energy distribution (SED) of TW Hya to models of SEDs that grains in the inner $0.5$ AU of the disk may have grown in
size and settled towards the midplane, while dust and gas in the outer disk 
is well-mixed.  The disk mass is roughly
$1.5-3 \times 10^{-2} M_\odot$ \citep{Wil00,Thi01,Tri01}. Based on
submillimeter CO emission and estimates for the disk mass, \citet{Thi01}
and \citet{Zad01} 
determined that CO is underabundant by a factor of $\sim500$ around TW Hya, 
with respect to a typical ISM H$_2$/CO ratio of $10^4$.
There is no molecular cloud associated with TW Hya
\citep{Web99}, and therefore no opportunity to observe whether or not
there is an outflow from H$_2$ or CO observations.  \citet*{Wei00} reported the detection of
H$_2$ emission in the 1-0 S(1) line at 2.12183 $\mu$m from TW Hya,
attributing this line to formation in the circumstellar disk, in which
the H$_2$ is excited by non-thermal electrons produced by X-ray ionization.
TW Hya is a strong X-ray source, with $L_x\sim
4\times 10^{-4}$ $L_\odot$ as measured with Chandra by
\citet{Kas01}. These Chandra observations indicated
an overabundance of Ne, an underabundance of O and Fe, and
solar-like abundances of Si, Mg and N in the X-ray emission region.  From
the sharply peaked emission measure distribution and high electron density, 
\citet{Kas01} concluded that the X-ray emission is produced in an accretion
shock.  Alternatively, \citet{Gag02} argued that the Chandra X-ray
emission may be consistent with an origin either in the accretion shock or
in a corona.

Here, we use \HST/STIS\ and \FUSE\ spectra of TW Hya 
to study its circumstellar 
environment, focusing on the substantial number of H$_2$
emission lines seen in the UV.  In \S\ref{section:observations} we explain the observations and data
reduction.  In \S\ref{section:analysis} we identify the emission lines and
examine their profiles.  In \S4 we show that the Si lines are anonomously
weak, and propose a depletion of Si in the accretion column as one possible 
explanation.  In \S5, we discuss the H$_2$ fluorescence, and in \S6 we
argue that this emission most likely originates in the circumstellar disk
of TW Hya.
In paper II, we model the H$_2$ line fluxes to assess the physical
conditions that give rise to the H$_2$ fluorescence.

\section{OBSERVATIONS AND DATA REDUCTION}
\label{section:observations}

We used the Space Telescope Imaging Spectrograph (\STIS) on Hubble
Space Telescope (\HST) to observe the pre-main sequence star TW Hya on
2000 May 7.  We used the medium resolution E140M grating and the
0\farcs5 $\times$ 0\farcs5 aperture for a 2300 s integration (see Table
\ref{tab:obs.tab}).  In this paper we only briefly discuss a
contemporaneous E230M spectrum of TW Hya.

We reduced the data using the CALSTIS software package \citep{Lin99}
written in the Interactive Data Language (\IDL), which
yields equivalent results to the HST archive pipeline.  Wavelengths were
assigned by the reduction software using calibration lamp spectra
obtained during the observations.  
A canonical wavelength solution is determined
from a 2-D fit of emission lines in a deep lamp exposure. Then a 
2-D cross-correlation of the deep exposure with a short lamp
exposure obtained during the TW Hya observations provides the
offset that must be applied to the canonical wavelength solution.
Absolute wavelengths are
accurate to $\sim 3$ \kms, and relative wavelengths are accurate to
$\sim 1$ \kms\ \citep{Lei01}.  Scattered light was removed from the
data using the ECHELLE\_SCAT routine available in the CALSTIS package,
which is identical to the {\it sc2d} algorithm used in the standard \HST\ pipeline.

We used the \FUSE\ satellite to observe TW Hya on 2000 June 6 with the LWRS
($30\arcsec \times 30 \arcsec$) aperture for a 2081 s integration
\citep{Her01}.  We reduced the data using v1.8 of the CALFUSE pipeline
\citep*{Kru01}.  \FUSE\ consists of 4 gratings with two channels each.  The 
spectrum was re-binned to one wavelength bin per resolution
element to increase signal-to-noise.
In this paper we use spectra from the SiC1A, LiF1A, LiF2B, and LiF2A channels,
covering the spectral regions 915--1006 \AA, 990--1080 \AA, 1095--1185, and
1090--1180 \AA, respectively.  We calibrate the wavelengths of the LiF1A
and LiF2A segments using H$_2$ lines, and of the SiC2A segment using
\ion{C}{3} ISM absorption, to obtain an accuracy of $\sim 0.1$ \AA.

\section{EMISSION LINES}
\label{section:analysis}
\subsection{Line Identification}

We detect in the \STIS\ E140M observation of TW Hya over 200 emission lines
in the 1140-1710 \AA\ spectral region, shown in Figures 1--2.  We use a database of
over 19,000 H$_2$ Lyman-band lines \citep{Abg93} and the \citet{Kur95}
atomic line database to identify lines based upon both coincidence with
calculated (for H$_2$) or lab rest wavelengths (for atoms) and consistency
of fluxes with predicted branching ratios from the upper state for H$_2$
lines.  

We
identify 143 H$_2$ lines, including five blends, from 19 excited
levels, listed in Table 2 by upper level.
Following conventional spectroscopic
notation, these lines are identified as R or P depending on the change in
rotational quantum number of $\Jup-\Jlo=-1$ or $+1$, respectively, where
$\Jup$ refers to the rotational quantum number of H$_2$ in the upper excitation state and
$\Jlo$ refers to the rotational quantum number in the lower excitation state.  For
example, the line 0-4 P(1) at 1335.921 \AA\ is the transition between
$\Jlo=1$, $\Vlo=4$ in the ground electronic state and $\Jup=0$, $\Vup=0$
in the excited electronic state.  We use the term {\it progression} to
indicate the set of R and P transitions from a single excited level to
the various available levels in the ground electronic state, as
demonstrated in Figure 3.
We
describe our identification of lines by analyzing in detail the lines from the $\Vup=0,
\Jup=17$ level in the excited electronic state.  Table
\ref{tab:vu0ju17.tab} shows the possible radiative transitions from
this state.  We detect five of the six transitions with the
largest radiative de-excitation rates, $A_{ul}$, and calculate that the
flux ratios are consistent with the branching ratios.  
The only undetected line among the strongest six transitions
is 0-5 P(18) at 1548.146 \AA.  The flux at this wavelength is dominated
by the blue side of the \ion{C}{4} 1548 \AA\ line, making the detection of
the H$_2$ line difficult.  In addition to these 143 H$_2$ lines from our
STIS observations, Figure 4 shows the detection of the
 Lyman-band H$_2$ transitions 1-1 P(5) at
1161.864 \AA\ and 1-1 R(3) at 1148.751 \AA\ in the \FUSE\ spectra of
TW Hya \citep{Her01}.

In Table \ref{tab:tentative} we list 15 tentatively identified
 H$_2$ Lyman-band lines from 10 different excited levels.
Based on branching ratios, these lines should be among the strongest
lines from a given excited level, and a transition to that excited
level coincides with a strong emission feature.  For these tentative
identifications, emission in the other transitions from the same upper
level is either masked by
strong emission features, occurs shortward of the spectral region we
observed, is obscured by a wind absorption feature in an atomic line, or is too weak to be clearly
identified as emission.  For example, we identify a narrow feature at
1593.826 \AA\
tentatively as emission in the H$_2$ transition 3-10 R(5), the second
strongest transition from $\Vup=3, \Jup=6$, with a
calculated rest wavelength of 1593.751 \AA.  
The strongest transition
from this excited state, 3-10 P(7) at 1611.315 \AA, appears to be
weakly present.  The third and fourth strongest lines, 3-1 R(5) and 3-1
P(7), have wavelengths shortward of the \STIS\ bandpass.  This observation is
too short to detect the weaker lines.  The excited level for this
transition may be pumped by \ion{C}{3} via the 3-2 R(5) transition at
1175.248 \AA.  We do not use these tentatively identified H$_2$ lines
in further analysis.

A similar test of over 75,000 CO lines from the Kurucz database failed
to identify any groups of CO lines in the spectrum of TW Hya; we
therefore consider the coincidence of individual CO transitions with
observed emission lines as random chance.  The usual electronic pumping
routes through the \ion{O}{1} multiplet at 1305 \AA, as seen in giant
stars such as $\alpha$ Boo \citep{Ayr86,Ayr01} do not appear to occur in TW Hya
(see Figure 5).  A number of
weak, narrow emission lines from 1354--1357 \AA\ may be CO A-X emission,
pumped by H$_2$ emission at 1393.9 \AA\ via transitions within the
thermal width of the H$_2$ line.

\subsection{Spatial Extent of Emission}

Observing with a $0\farcs5 \times 0\farcs5$ aperture, for TW Hya's distance 
of 56 pc, we
detect H$_2$ emission that occurs spatially within 14 AU of the star.
The \STIS\ E140M mode can provide some information on the spatial extent of
the H$_2$ emission across the aperture.  
We analyze separately the spatial distribution of H$_2$ lines with wavelengths $\lambda<1350$ \AA, $1350<\lambda<1500$ \AA, and  
$\lambda>1500$ \AA,
 because the width of the instrumental spatial profile increases
towards lower wavelengths \citep{Lei01}.  In this analysis, we do not use
H$_2$ lines contaminated by \ion{N}{5}, \ion{C}{2}, \ion{Si}{4}, or
\ion{C}{4} emission. We co-add the spatial extent of many H$_2$ lines and
compare their
spatial distribution to
other lines with similar wavelengths. 
The \Lya\ 1215.67 \AA\ line,
the \ion{C}{4} 1549 \AA\ doublet, and the \ion{He}{2} 1640 \AA\ line are
probably produced in the accretion shock near the stellar surface
\citep[e.g.,][]{Cal96} and therefore should be
spaitially unresolved and represent a good estimate of the spatial profile in the cross-dispersion direction.
Cooler lines, such as the \ion{O}{1} 1304 \AA\ triplet and the
\ion{C}{1} multiplets at 1561 \AA\ and 1657 \AA, may be produced in the
accretion shock or may also be somewhat extended.  The geocoronal \Lya\
line uniformly fills the aperture.  

We use the order locations on the detector to identify the location 
of each H$_2$ line and then extract a subimage, a 
$15\times15$ pixel region around the brightest pixel at this location.
We co-add these subimages, and then create normalized spatial profiles for H$_2$ from 3 pixel (0\farcs11)
wide swaths and for atomic lines from 7 pixel 
(0\farcs26) wide swaths centered on the 
brightest pixel.  Although the 
brightest pixel is offset in different wavelength bins by up to half a pixel,
 this only slightly affects the 
central region of the spatial profile and does not affect
the wings of the profile.  
Figure 6 shows that the H$_2$ profile has more power in
the wings of the spatial profile than the high-temperature lines and
\ion{O}{1}, although the wings are comparable to the extent of \ion{C}{1}
emission.  Geocoronal \Lya\ emission, which uniformly fills the aperture, is clearly more extended than the
H$_2$ emission.  

In Figure 6c we compare the spatial extent of
\ion{C}{4} emission and H$_2$ emission lines above 1500 \AA.  We split
these H$_2$ lines into 9 lines with flux above and 36 lines below $3\times10^{-14}$ erg
cm$^{-2}$ s$^{-1}$.  The stronger H$_2$ lines do not show the broad wings
that the weak emission lines show.  Similarly, we compare \ion{C}{4}
emission at the peak wavelength to \ion{C}{4} emission on
the edge of its spectral profile.  The spatial profile from the peak
emission strongly resembles the spatial profile of the strong H$_2$ lines,
while the spatial profile from the weak wings of the \ion{C}{4} profile
strongly resembles the spatial profile of the weak H$_2$ emission lines.
This relationship, with weak H$_2$ lines having broader wings than strong
H$_2$ lines, holds for all wavelengths.  This weak emission in the wings may be due to
a weak extended background UV source or detector background.
We conclude that the H$_2$ emission is not extended in the cross-dispersion 
direction beyond the $0\farcs05$
resolution of the telescope, comparable to 2.8 AU, or a 
1.4 AU radius from the 
central star.

\subsection{Line Profiles}
\label{section:velocity}
We fitted most of the emission lines with either a single or, for blended
lines, multiple Gaussians.  Several lines were not fitted because their
profiles are non-Gaussian, as shown by many examples in Figure 7.  The profiles of \Lya\ 1215.67 \AA, the
\ion{C}{4} 1549 \AA\ doublet, the \ion{N}{5} 1240 \AA\ doublet, the
\ion{He}{2} 1640.5 \AA\ line, the \ion{C}{3} 977 \AA\ line, and the
\ion{C}{2} 1335.7 \AA\ line show strong redshifted emission with a
sharp blue edge.  ISM absorption appears in the \ion{O}{1}
1302 \AA, \ion{C}{2} 1334.5 \AA\ line, \ion{Si}{2} 1526.7 \AA, and in the \ion{Mg}{2} 2800 \AA\
doublet, centered at the rest wavelength of these transitions in the
heliocentric frame.
The
low-temperature \ion{O}{1} 1305 \AA\ triplet, \ion{C}{2} 1335 \AA\ doublet, \ion{N}{1} 1492.6 \AA, and \ion{Si}{2} lines at 1526.7 \AA\ and
1533.4 \AA, along with the \ion{Mg}{2} 2800 \AA\ doublet in the NUV,
have P-Cygni profiles that indicate winds of $\sim230$ \kms.  Physical
properties that can be inferred from these lines will be determined 
in a subsequent paper.

We do not observe H$_2$ absorption within the \Lya\ profile, even though
the H$_2$ must absorb the \Lya\ photons to cause the fluorescence (see
\S5).  In \HST/STIS\ observations of Mira B, \citet*{Woo02} detect H$_2$
absorption against \Lya\, as well as the corresponding fluorescence.  Our
\FUSE\ observations do not show significant H$_2$ absorption against the
\ion{O}{6} profiles.  These unseen H$_2$
lines, with $\Vlo=0, \Jlo=1-4$, have been observed in other PMS stars
\citep{Rob01,Her01}.  Because there is no H$_2$ absorption in our line of
sight, if the H$_2$ is in a 
circumstellar shell or outflow around TW Hya, then the shell or outflow must be
clumpy enough such that the H$_2$ is not in our line of sight to the star.
In Paper II we will quantify this conclusion.

The observed \Lya\ profile is characterized by a strong red wing, possibly
produced in the accretion shock, and a dark, wide absorption feature that
extends to -700 \kms.  The large width of the \Lya\ line may be caused 
by pressure broadening, which \citet*{Muz01} argue produces the broad
wings of H$\alpha$ based on modelling its line profile.
  The absorption feature is probably caused by a combination of wind, 
circumstellar, and interstellar absorption.  In Paper II we will analyze the 
\Lya\ profile in greater detail, including on a comparison between the
observed \Lya\ profile and a synthetic \Lya\ profile reconstructed from
H$_2$ emission line fluxes.

To correct for instrumental line broadening, we now calculate the spectral
line-spread function of the detector.  We determined in \S3.2 that emission
in both atomic and molecular lines is not spatially extended in the
cross-dispersion direction.  If the emission is also not
extended in the dispersion direction, then from the spatial distribution of
emission we can determine the spectral line-spread function.  We fit
multiple Gaussians to the spatial distribution of H$_2$ in three regions:
$\lambda<1350$, $1350<\lambda<1500$, and $\lambda>1500$.  We then multiply
the FWHM of these Gaussians by 0.82, which is the ratio of the plate scale
of the detector in the cross-dispersion direction to the plate scale in the 
dispersion direction due to anamorphic magnification.  We then fit emission lines with Gaussians,
correcting for the spectral line-spread function, to obtain intrinsic line
widths.  Parameters of these fits to atomic and
unidentified lines are listed in Table
\ref{tab:linelist.tab}.  We only list the flux from our Gaussian fits to
H$_2$ lines because there are so many and they are all similar.  We
measured the flux of the two \FUSE\ H$_2$ in the LiF1B and LiF2A channels
separately and co-added the flux of each line, weighted by the error in the 
flux.

The $0\farcs5 \times 0\farcs5$ aperture is not a supported operating mode
for STIS.
The line-spread function we calculate from the spatial profile of the H$_2$ 
emission is similar to the line-spread function determined by \citet{Lei01} 
for the $0\farcs2 \times 0\farcs2$ aperture,
and as expected, they are slightly broader for the  $0\farcs5 \times
0\farcs5$ aperture.
We show in Figure 8 that the mean observed FWHM of H$_2$ lines is $18.16
\pm 0.10$\footnote[2]{Errors are $1-\sigma$ errors throughout paper}
\kms, before removing instrumental broadening.  Two lines, 1-8
P(11) at 1603.164 \AA\ and 1-4 R(3) at 1314.690 \AA, have anomalous
widths of more than 30 \kms\ and are not used in future analysis. 
After
removing instrumental broadening with our calculated spectral line-spread
function, the average FWHM of the 30 H$_2$ lines with
fluxes above $2\times10^{-14}$ erg cm$^{-2}$ s$^{-1}$ is $11.91 \pm 0.16$ 
\kms, 
which corresponds to a temperature of 6200 K if the broadening were
entirely thermal.  Although we believe our calculated spectral
line-spread function is more accurate for our data than the line-spread function given
in \citet{Lei01} for the $0\farcs2 \times 0\farcs2$ aperture, we calculate
a FWHM of $14.2\pm0.2$ \kms\ when we correct for instrumental broadening using
the line-spread function for the $0\farcs2 \times 0\farcs2$ aperture.
In Paper II we will determine the temperature of the
H$_2$ emission region and calculate the non-thermal broadening of the H$_2$ 
emission.
%

Figure 9 shows that the mean velocity shift of 137 H$_2$ lines is $13.55 \pm 0.10$ \kms, with respect to rest wavelengths calculated by \citet{Abg93}.  We used a conservative error estimate of 0.75 \kms\ for the assumed random uncertainty in the
rest wavelengths, which is slightly larger than the difference between
observed and calculated
wavelengths, as given by \citet{Abg93}.  The velocity shift of H$_2$ lines,
given the approximate $\pm 3$ \kms\ uncertainty in wavelength 
zero-point calibration
for STIS E140M spectra \citep{Lei01}, is
consistent with previous photospheric radial velocity measurements:
$12.2\pm0.5$ \kms\ \citep{Wei00}, $12.2\pm0.1$ \kms\ \citep{Kas99},
$12.5\pm2.2$ \kms\ \citep{Del89}, and $12.9\pm0.2$
\kms\ \citep{Tor01}.  The velocity shift for 30 lines shortward of 1350 \AA\ is $13.4\pm0.2$
\kms, compared with $13.3\pm0.1$ \kms\ for 67 lines between 1350
\AA\ and 1550 \AA, and $14.2\pm0.2$ \kms\ for 29 lines above 1550 \AA,
confirming the excellent relative wavelength scale of \STIS. 

\section{HOT ACCRETION LINES}
\label{section:discussion}

Emission in lines of \ion{C}{4}, \ion{Si}{4}, \ion{N}{5}, and
\ion{O}{6}, from gas at $\sim10^5$ K, is probably produced in an accretion flow near the surface of the
star \citep{Cal96,Lam98}, as indicated by their broad red wings, which may be
produced by rapidly inflowing hot gas.  This emission may also be produced
in a stellar transition region, although these red wings are broader than
emission from most active stars with a coronal emission source
\citep[e.g.,][]{Ayr01a}.  The ratio of flux in the
\ion{C}{4} doublet near 1550 \AA\ to the flux in the \ion{Si}{4}
doublet at 1400 \AA\ is 30.  In a study of eight CTTSs with \HST/GHRS,
\citet{Ard01} found that this ratio varies from 0.5 to 3 in stars that show
\ion{Si}{4} emission, but \ion{Si}{4} emission was conspicuously absent
from a number of these stars.
In spectra of 
main-sequence and giant stars, the 
flux ratio of these two doublets is typically about two
\citep[e.g.,][]{Ayr97}.  In giant stars such as $\alpha$ Boo, the Si IV
doublet may be absorbed by \ion{Si}{1} bound-free absorption below 1525
\AA\ \citep{Ayr01}.  When this absorption occurs, we expect \ion{C}{1} continuum absorption to
quench any \ion{N}{5} emission, as is observed for $\alpha$ Boo (Ayres,
private communication), but we detect strong \ion{N}{5}
emission at 1238 \AA\ and 1243 \AA.  Temperature effects probably do
not explain the weakness of the \ion{Si}{4} features because it is formed at
80,000 K, close to the 100,000 K formation temperature of \ion{C}{4}.
Strong emission in the red wing of the \ion{C}{3} line at 977.0 \AA\
indicates the presence of cooler material in
the accretion flow, while a strong red wing in \ion{C}{4}, \ion{N}{5}, and \ion{O}{6}
emission indicates hotter material in the flow.

We do not detect \ion{Si}{3} features at 1206 \AA\ and 1296 \AA, which
are strong in main-sequence and active stars, and the \ion{Si}{2}
features at 1309.3 \AA, 1526.7 \AA, and 1533.3 \AA\ are weak.  In addition, 
\citet{Val00} find in IUE spectra of TW Hya that the \ion{Si}{3}] 1892
\AA\ line is also fairly weak, although a reasonable amount of flux is present
in the \ion{Si}{2}] 1808 \AA\ and \ion{Si}{2}] 1817 \AA\ lines.
 All of this leads us to speculate that Si may be underabundant in the accretion
flow.  In the gas phase of the ISM, Si is underabundant
 by a factor of 100
relative to solar because Si readily depletes onto grains
\citep{Fit96}.  \citet{Chi01} and \citet{Dal99} provide evidence
that grains in a circumstellar disk settle near the midplane.
  \citet{Gam96} showed that
ionized material at the surface of the disk couples to the magnetic
field, and thereby preferentially accretes onto the star.  Silicon may
therefore be underabundant in the accretion flow because it is
predominantly located in grains at the circumstellar midplane that do
not participate in the accretion.  If most of the Si is bound in
grains, then we might expect strong SiO emission from stellar disks when
the accretion shock has underabundant \ion{Si}{4} emission, since the
formation of SiO is one step towards the incorporation of Si into dust.
\citet{Wei01} and \citet{Sit00} find a broad hump in the IR region 8-13
$\mu$m, which they attribute to a combination of silicates, and
\citet{Wei01} conclude that the disk of TW Hya contains several hundred
Earth masses of condensed silicates and ices.
If
depletion onto grains in a TTS environment resembles similar processes
in the ISM, then other refractory elements such as Fe should also be underabundant.
Unfortunately, analysis of the strength of NUV \ion{Fe}{2} lines is
complicated, partly
because most of them are photoexcited by \Lya\ \citep*{Car94,Mcm00}, and
there are no other Fe lines available for us to study.
\citet{Kas01} found solar-like abundances of Si and Mg, an underabundance 
of Fe and O, and an overabundance of Ne in the Chandra spectrum of
TW Hya.  They attributed the X-ray emission to the accretion shock, which
disagrees with previous models in which X-rays from TTSs are produced
in a coronae \citep[e.g.,][]{Fei99}.  If this is the case, then our
speculation that Si is underabundant in the accreting gas would be
incorrect.
  We also note that we do not detect
\ion{Cl}{1} emission at 1351.7 \AA, which is strong in most stars due to
pumping by \ion{C}{2}.  We will pursue these abundance issues further in a
future paper.

\section{H$_2$ FLUORESCENCE}
\label{section:results}

We identify 19 progressions of H$_2$ lines, each with a distinct upper
vibrational level $\Vup$ and rotational level \Jup.  In Table
6, we list these progressions, including the number
of detected lines and flux ($F_{\rm obs}$) in each progression.
Ten other progressions listed in Table 3 include
the 15 tentatively identified H$_2$ lines.  Figure 10
shows that most levels in the excited electronic state are photoexcited
by transitions coincident with \Lya.  Figure 11 shows that the lower levels
of these transitions are the lowest energy levels with transitions
coincident with \Lya, indicating that the level populations may be thermal.
The \ion{C}{4} doublet,
\ion{C}{2} doublet, and \ion{C}{3} multiplet also pump some of these
levels.

The observed \Lya\ profile includes a broad absorption feature with
FWHM about $500$ \kms, produced by neutral H in the ISM, circumstellar
material, and a wind from TW Hya or its disk.  We detect no flux due to
either \Lya\ or H$_2$ emission in this dark absorption line.  This absorption feature
coincides with several H$_2$ transitions that pump the excited state,
so most of the \Lya\ absorption must occur after the \Lya\ emission
encounters the H$_2$.

We demonstrate the identification of a pumping mechanism for a given
excited level by returning to the case of $\Vup=0, \Jup=17$.  
Because most of the H$_2$ is in the ground electronic state, and the low
transition probabilities preclude any collisional redistribution of H$_2$ 
within the excited electronic state,
one of the transitions shown in Table
\ref{tab:vu0ju17.tab} must pump the $\Vup=0, \Jup=17$ level.  
Of these
transitions, only the 0-5 P(18) 1548.213 \AA\ line overlaps with a
strong emission line observed in the TW Hya spectrum.  The weak H$_2$
feature at 1381 \AA\ is unable to pump a significant amount of H$_2$
into the $\Vup=0, \Jup=17$ level.  The continuum at the
wavelengths of these transitions is also unable to pump an appreciable
amount of H$_2$ into this level.  The 0-2 R(16) transition, which would be
observed at 1335.357 \AA, falls in between the \ion{C}{2} 1335 \AA\
doublet.  We do not expect significant blue emission from the 1335.7 \AA\
line in this doublet.  The \ion{C}{2} 1334.5 \AA\ line probably has a strong red
wing similar to the \ion{C}{2} 1335.7 \AA\ line, but this wing is absorbed
by a wind in \ion{C}{2} 1335.7 \AA.
  We estimate that the strength of the \ion{C}{2} 1334.5 \AA\ line at
+180 \kms\ is $\sim 10^{-13}$ erg cm$^{-2}$ s$^{-1}$, which is the strength of the \ion{C}{2} 1335.7 \AA\ line at the same position.
This flux is a factor of 10 weaker than the \ion{C}{4} flux available to
pump this upper level.  The lower levels for 0-2 R(16) and 
0-5 P(18) have energies
2.8 eV and 3.8 eV, respectively, that are too high to
have a significant population at sensible temperatures.  The 
$\Vlo=5,\Jlo=18$ level is particularly difficult to populate.  We conclude that
either the 0-5 P(18) transition at 1548.213 \AA\ or the 0-2 R(16)
transition at 1335.357 \AA\ pumps this upper level.  The pumping mechanism
for almost all other levels is well determined.

The pumping transitions listed in Table 6 require significant
populations in the lower rovibrational levels.  Changes in these level
populations may alter which pumping mechanism is most important for
certain excited levels.  At low temperatures, the ground vibrational level will be much
more populated than excited vibrational levels.  As a result, $\Vup=0,
\Jup=0$ and $\Vup=0, \Jup=3$ would be pumped by the continuum via  0-0
P(1) at 1110.062 \AA\ and 0-0  P(2) at 1110.119 \AA, respectively.
However, because we do not observe other progressions that would be
pumped by the continuum, we conclude that the gas temperature is
sufficiently high that \Lya\ pumps these excited levels.  The $\Vup=3,
\Jup=3$ level is pumped by both \ion{C}{3} 1174 \AA\ via the 3-2 P(4)
transition at 1174.923 \AA\ and by \Lya\ via the 3-3 P(2) transition at
1217.031 \AA.  Pumping
of the $\Vup=0, \Jup=17$ and the tentative route $\Vup=0, \Jup=24$ cannot
possibly occur if the H$_2$ populations are thermal because the energy of
the lower
level for the routes which populate these levels are very high; we
claim that the \ion{C}{4} emission pumps these former via 0-3 P(25) at
1547.971 \AA, and that \ion{C}{4} and \ion{C}{2} emission pumps the latter, 
as described above.  The lower level $\Vlo=3,\Jlo=25$ for the 
first transition has an energy of 4.2 eV, 
compared with the 4.5 eV dissociation energy of the H$_2$
molecule.

One
H$_2$ transition, 0-4 R(1) at 1333.851 \AA, with a flux of 
$(7.9\pm0.7)\times10^{-15}$ erg cm$^{-2}$
s$^{-1}$, is about six times weaker than would be expected by branching
ratios.  Its counterpart, 0-4 P(3) at 1342.314 \AA, has a similar branching 
ratio and lower energy level, and a flux of
$(64.9\pm1.3)\times10^{-15}$ erg cm$^{-2}$.
In Paper II, we will
show quantitatively that this particular line is anomalously weak,
and that it is the only line that is signifcantly weaker than predicted by
our modelling.  We conclude that this H$_2$ line, which is centered at 130
\kms\ to the blue side of the \ion{C}{2} 
1334.5 \AA\ line, is absorbed by a wind
component of this \ion{C}{2} line.  This result places the wind in our line 
of sight to the 
H$_2$ emission region, and constrains the plausible geometries of TW Hya's
circumstellar environment.

We now analyze the velocity shifts and line FWHMs pumped from levels with
$\Elo<1.25$ eV and $\Elo>1.5$ eV separately.  As discussed in \S3.3, the
velocity shift of H$_2$ lines increases slightly with increasing
wavelength, probably because of the relative wavelength accuracy of \STIS.
We correct for these changes by fitting a line to the velocity shift and
wavelength of each transition.  We find that 
the velocity shift of
37 lines 
pumped from high
energy levels is an average of $0.70 \pm 0.16$ \kms\ above the
estimated velocity, while the velocity shift of the 65 lines pumped from
low-excitation levels is $-0.94 \pm 0.15$ \kms\ from the estimated velocity
shift.  The accuracy of the wavelengths calculated by
\citet{Abg93} compared with laboratory wavelengths indicates that there
is most likely not a wavelength bias for high-energy lines.  Thus,
the difference in the mean velocity shift between the two groups of
H$_2$ lines is likely real.

\section{ORIGIN OF H$_2$ EMISSION}
\label{section:disk}
We now consider a circumstellar shell, an outflow, the stellar photosphere, and a disk as 
possible places of origin for the H$_2$ emission.  We have a number of
pieces of evidence to discriminate between models.  The H$_2$ emission is
not shifted by more than 3 \kms\ relative to the photospheric radial
velocity.
The spatial extent of the emission indicates that most of it
occurs within $0\farcs05$ from the star.  The \Lya\ profile shows no H$_2$
absorption, which gives an upper limit to the amount of H$_2$ in our line
of sight.  Finally, the \ion{C}{2} wind feature absorbs H$_2$ emission,
which indicates that \ion{C}{2} lies in our line of sight to the warm H$_2$.

Warm molecular emission has in the past been interpreted as shock-heated
emission due to an outflow interacting with the surrounding ambient molecular
material.  However, the TWA is isolated from any molecular cloud, so this
type of emission is difficult to accept.  Molecular emission is absorbed
by a wind component in the \ion{C}{2} line at 1334.5 \AA, which would not
occur in most shell models.  The H$_2$ progressions pumped near
line-center of Ly$\alpha$ probably would not occur because of \ion{H}{1}
absorption in the wind.  Assuming the outflow interacts with
surrounding material at a large distance from the star, the H$_2$ emission
should be more extended.  Consequently, we
rule out emission from a surrounding molecular cloud.

Molecular gas could become entrained in an outflow, or perhaps form
part of the outflow, which could produce
emission from hot H$_2$ gas.  In this scenario, the star would have to be
extremely close to face-on so that we do not observe either spatial extent
in the H$_2$ emission or H$_2$ absorption against the Ly$\alpha$ emission.
However, we rule out velocity shifts greater than 3 \kms\ for the H$_2$
emission, which is inconsistent with expected outflow velocities.

\citet{Joh01} determined that CO absorption near 1.6 $\mu$m occurs in the
photosphere of most TTSs.
The data analyzed here are roughly consistent with an
origin of the H$_2$ emission
in the photosphere.  If the
H$_2$ is produced in the photosphere, then it should be detected in the IR around
Naked TTSs (NTTS) as well as CTTSs, but this emission has not yet been 
detected.  In a study of H$_2$ fluorescence with \HST/GHRS, \citet{Ard01}
detected H$_2$ emission from eight
 CTTSs but did not detect H$_2$ emission from 
the one NTTS in their sample.  This result is consistent with a disk origin 
of H$_2$ emission, if the 
gas disappears from the disk on the same timescales that dust
disappears.  However, because this
H$_2$ emission is reprocessed \Lya\ emission, NTTSs may not show H$_2$
emission because their atomic emission lines are typically weak.  Even with 
weak profiles, \Lya\ may still be strong near the center of the line
profile. In Figure 10, two
progressions are pumped near the rest wavelength of \Lya, and fluoresence
in these progressions could in principle be seen from NTTSs.  This
comparison requires further observations with greater sensitivity than
\HST/GHRS offered in order to interpret the results.

We conclude that the most likely site of H$_2$ emission is in the disk.  A
disk origin satisfies all of our requirements listed above.  The fluorescence could only occur in the surface
layer of the disk, because grains would absorb any UV radiation below a
certain depth.  In Paper II, we will calculate the physical properties of
the H$_2$ emission region.  The H$_2$ fluorescence typically occurs in warm 
regions ($T\sim2000-3500$ K), which may pose certain problems in comparing
this conclusion with disk models \citep[e.g.,][]{Dal01a,Chi01,Gla01}.  From the
temperature of the disk, we will calculate the thermal broadening of the
H$_2$ emission and subsequently spatially constrain the H$_2$ emission
using its orbital velocity.  

If this H$_2$ fluorescence occurs in a disk, then the line profiles from
stars with higher inclinations should be wider.
In \HST/STIS spectra of
the DF Tau, with an edge-on disk, \citet*{JLL99} found that FWHM of the H$_2$ emission lines 
were typically about 25 \kms, while we calculate a FWHM of 18 \kms for the
face-on system TW
Hya.  \citet{Joh01} determined from CO absorption that the $v\sin i$ of DF Tau is $18.5 \pm 4$ \kms, compared
with $<6$ \kms for TW Hya.  The additional broadening observed in the
H$_2$ lines of DF Tau may result either from a faster orbital velocity if
the H$_2$ is in the disk, or a faster stellar rotation if the H$_2$ is in
the photosphere.


\section{SUMMARY}
\label{section:summary}

Our analysis of the ultraviolet spectra of TW Hya obtained with the STIS
instrument on \HST\ and with \FUSE\ lead to the following conclusions:

1.  The 1250--1650 \AA\ spectrum is dominated by molecular hydrogen
fluorescence in 143 Lyman band lines.  The observed flux in these lines
is $1.94 \times 10^{-12}$ erg cm$^{-2}$ s$^{-1}$, corresponding to
$1.90 \times 10^{-4} L_\odot$ at the stellar distance of 56 pc.  The
H$_2$ spectrum is pumped to the excited electronic state mainly by
transitions coincident with the broad \Lya\ emission line.  Other
lines, such as the \ion{C}{2} 1335 \AA\ doublet, \ion{C}{3} 1175
\AA\ multiplet, and the \ion{C}{4} 1550 \AA\ doublet may also pump some
H$_2$ to the excited electronic state.  

2.  The H$_2$ emission is not spatially extended from the star,
given a resolution of $0\farcs05$, corresponding to 2.8 AU at 56 pc.  With a face-on 
disk, the H$_2$ emission is produced within 1.5 AU of the central star.


3.  The wavelengths of the H$_2$ emission is not shifted by more than 3
\kms\ with respect to the photospheric velocity of the star.

4.  The stellar wind occurs in our line of sight to the H$_2$ emission
region, based upon the anomolously weak flux in an H$_2$ line that has a
wavelength coincident with a wind feature of
\ion{C}{2} 1334.5 \AA.

5.  No significant H$_2$ absorption is detected against the \Lya\ profile
or the \ion{O}{6} emission lines.

6.  The H$_2$ emission likely originates in the surface layer of a disk,
rather than a circumstellar shell or an outflow.  We cannot rule out the possibility that the H$_2$
emission is produced in the stellar photosphere.

7.  Silicon II, III, and IV lines are anomolously weak in the UV, possibly because Si depletes onto
grains that accumulate in the disk midplane and decouples from the
surface material of the disk that preferentially accretes.  Thus the
accretion column responsible for the high temperature emission lines would
then be
very silicon-poor.  Models of the accretion shock are needed to further
probe this possibility.

\section{ACKNOWLEDGEMENTS}

This research is supported by NASA grant S-56500-D to the University of
Colorado and NIST, and by a grant to the University of Colorado
 through the Johns Hopkins University.  We thank Tom Ayres, Phil
Maloney, David Hollenbach and Bruce
Draine for valuable discussions.


\begin{figure}
\resizebox{6.5in}{8.in}{\plotone{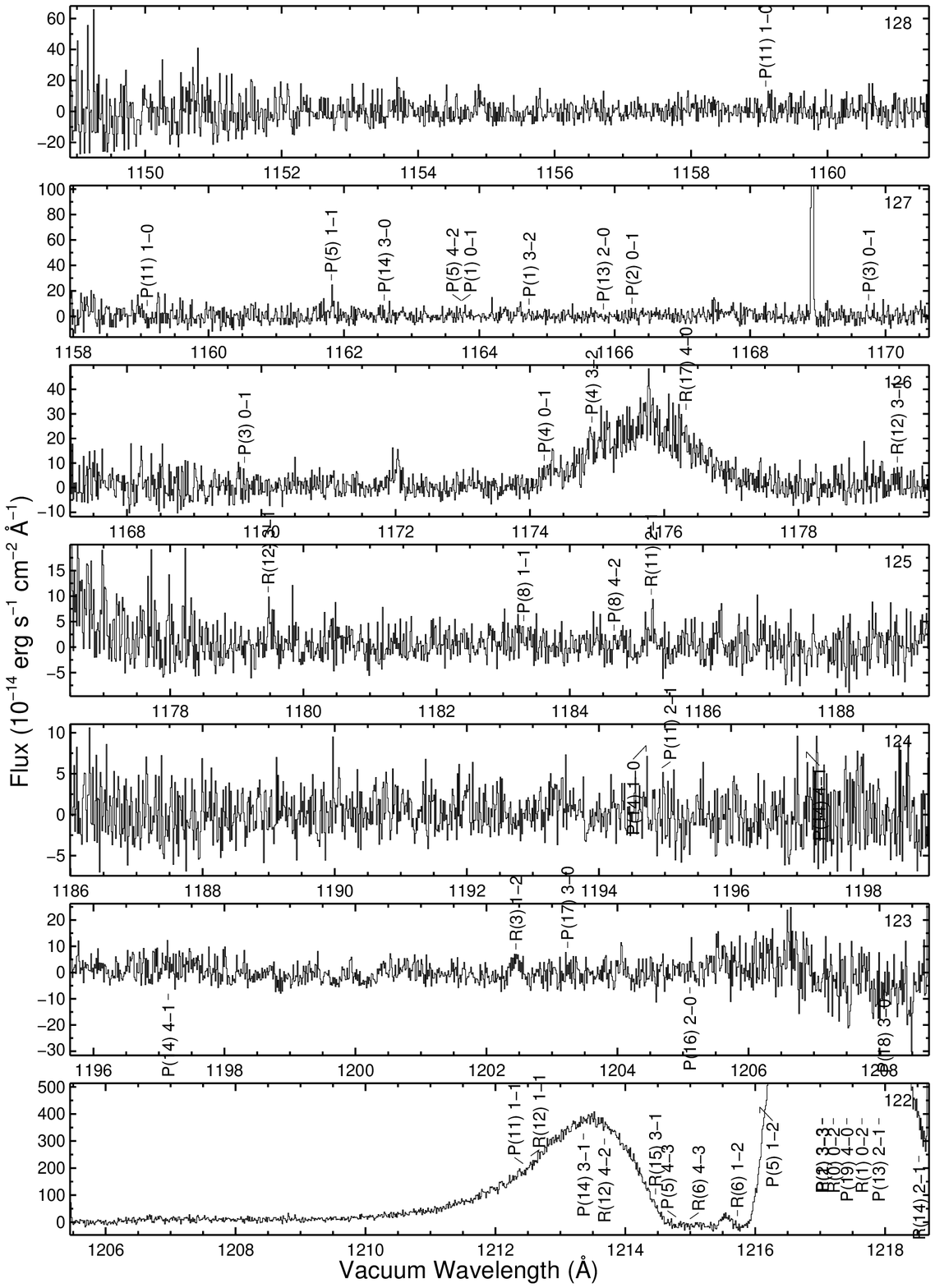}}
\caption{\textit{HST}/STIS E140M spectrum of TW Hya.  Lyman-band
transitions of H$_2$ that may be pumped by \Lya\ are labeled.  Some of
these transitions are too weak to be detected.}
\label{fig:herczeg1.eps}
\end{figure}

\setcounter{figure}{0}

\begin{figure}
\resizebox{6.5in}{8.in}{\plotone{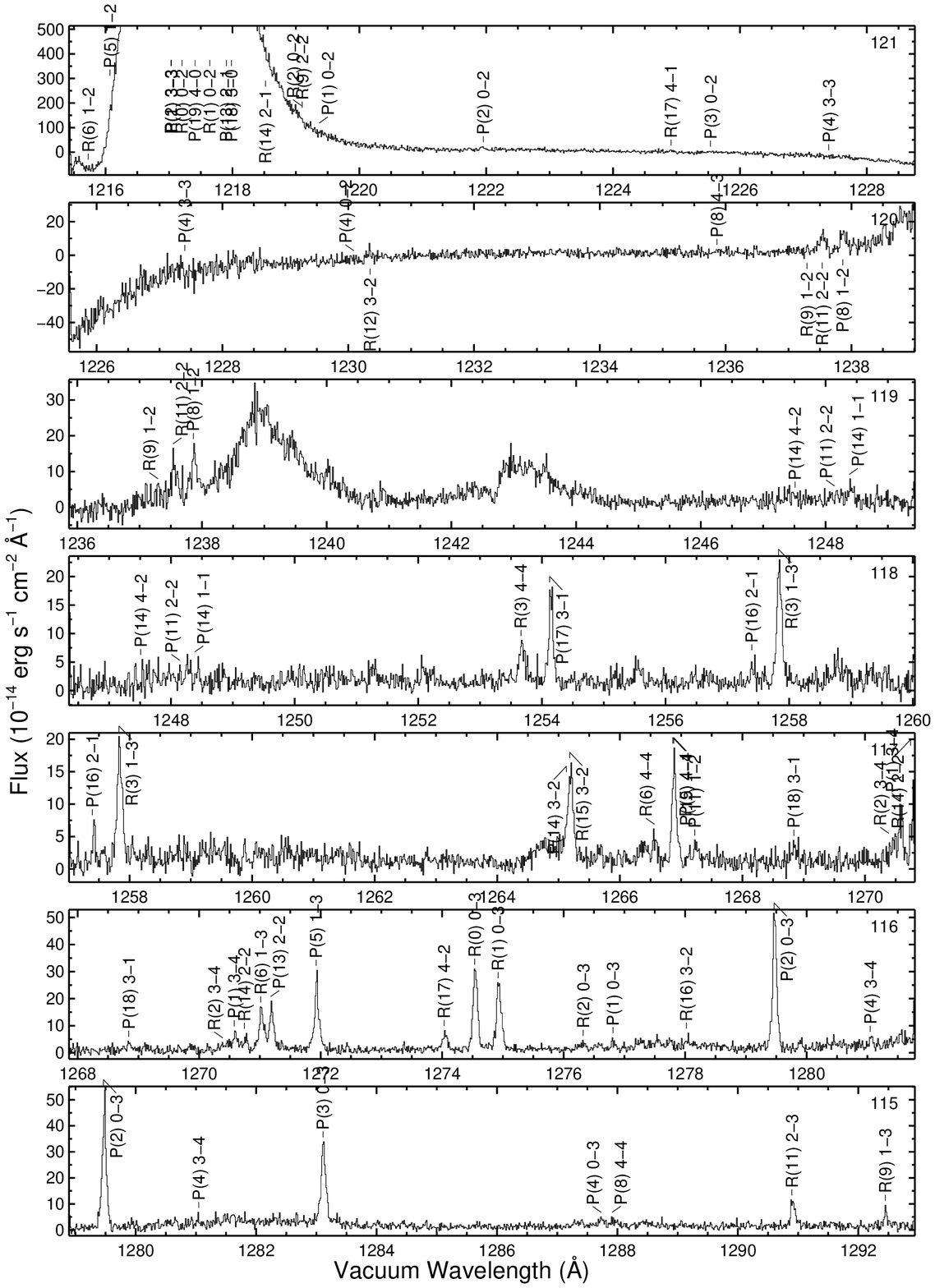}}
\caption{Continued}
\label{fig:herczeg2.eps}
\end{figure}
\setcounter{figure}{0}

\begin{figure}
\resizebox{6.5in}{8.in}{\plotone{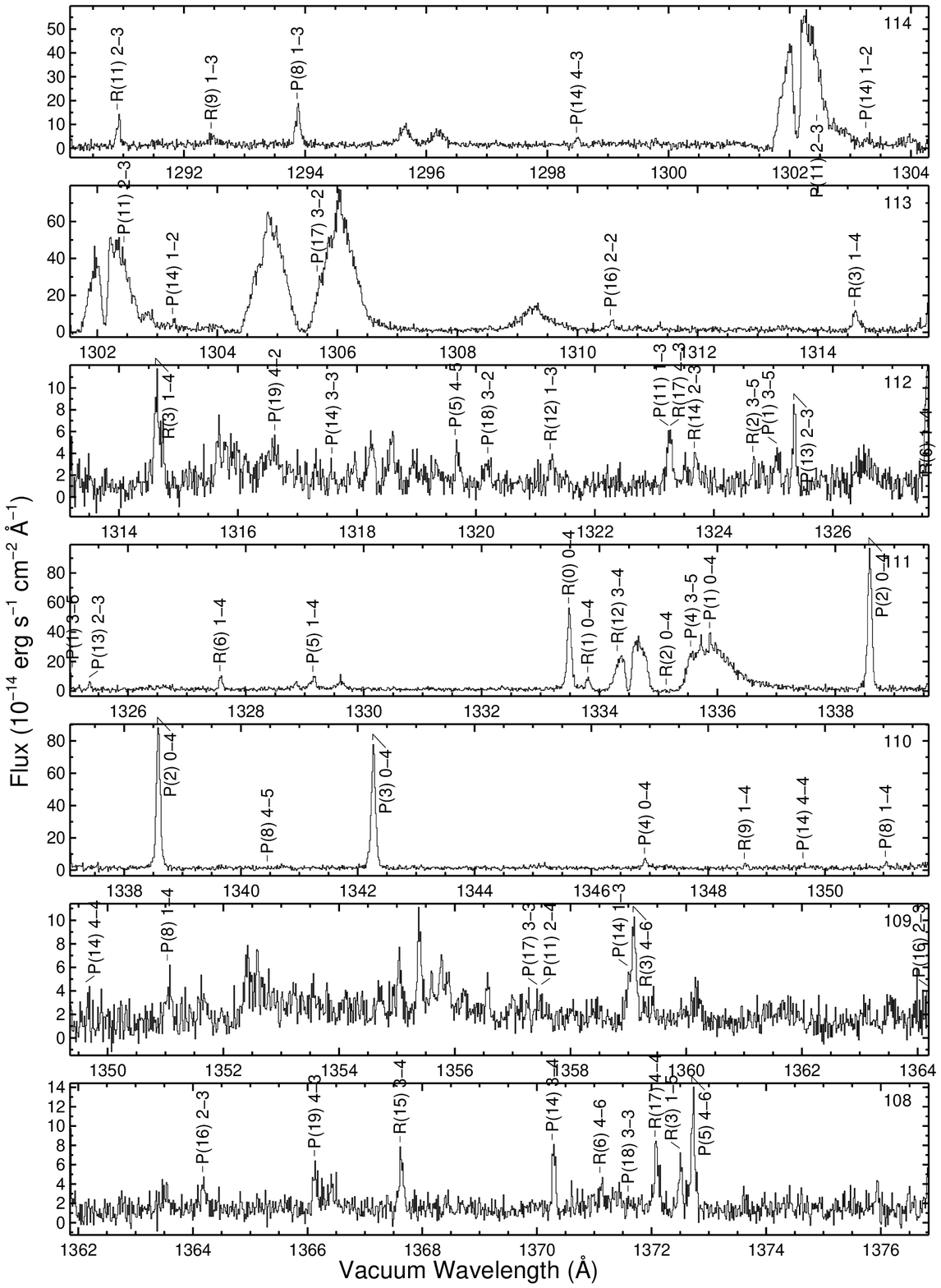}}
\caption{Continued}
\label{fig:herczeg3.eps}
\end{figure}
\setcounter{figure}{0}

\begin{figure}
\resizebox{6.5in}{8.in}{\plotone{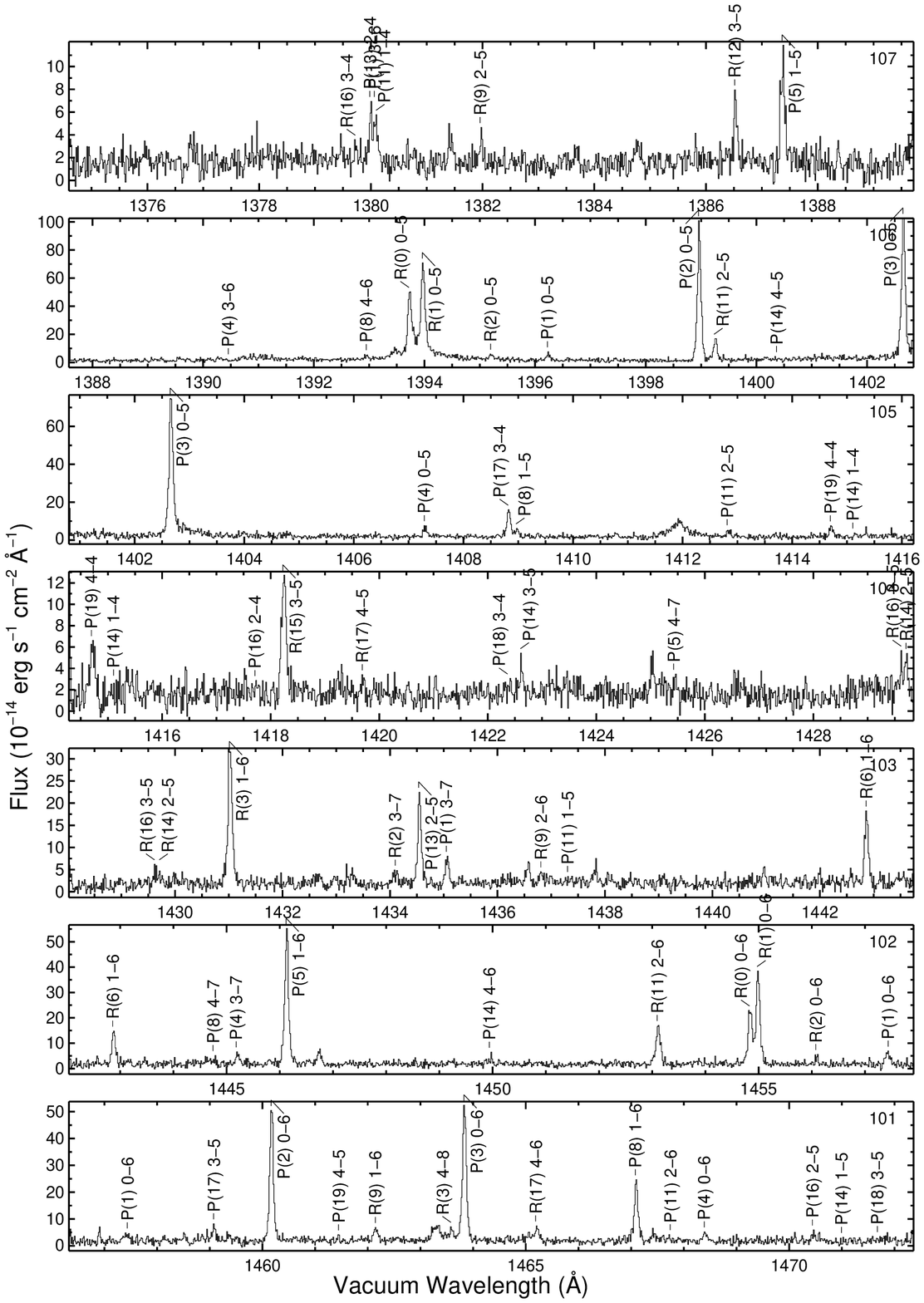}}
\caption{Continued}
\label{fig:herczeg4.eps}
\end{figure}
\setcounter{figure}{0}

\begin{figure}
\resizebox{6.5in}{8.in}{\plotone{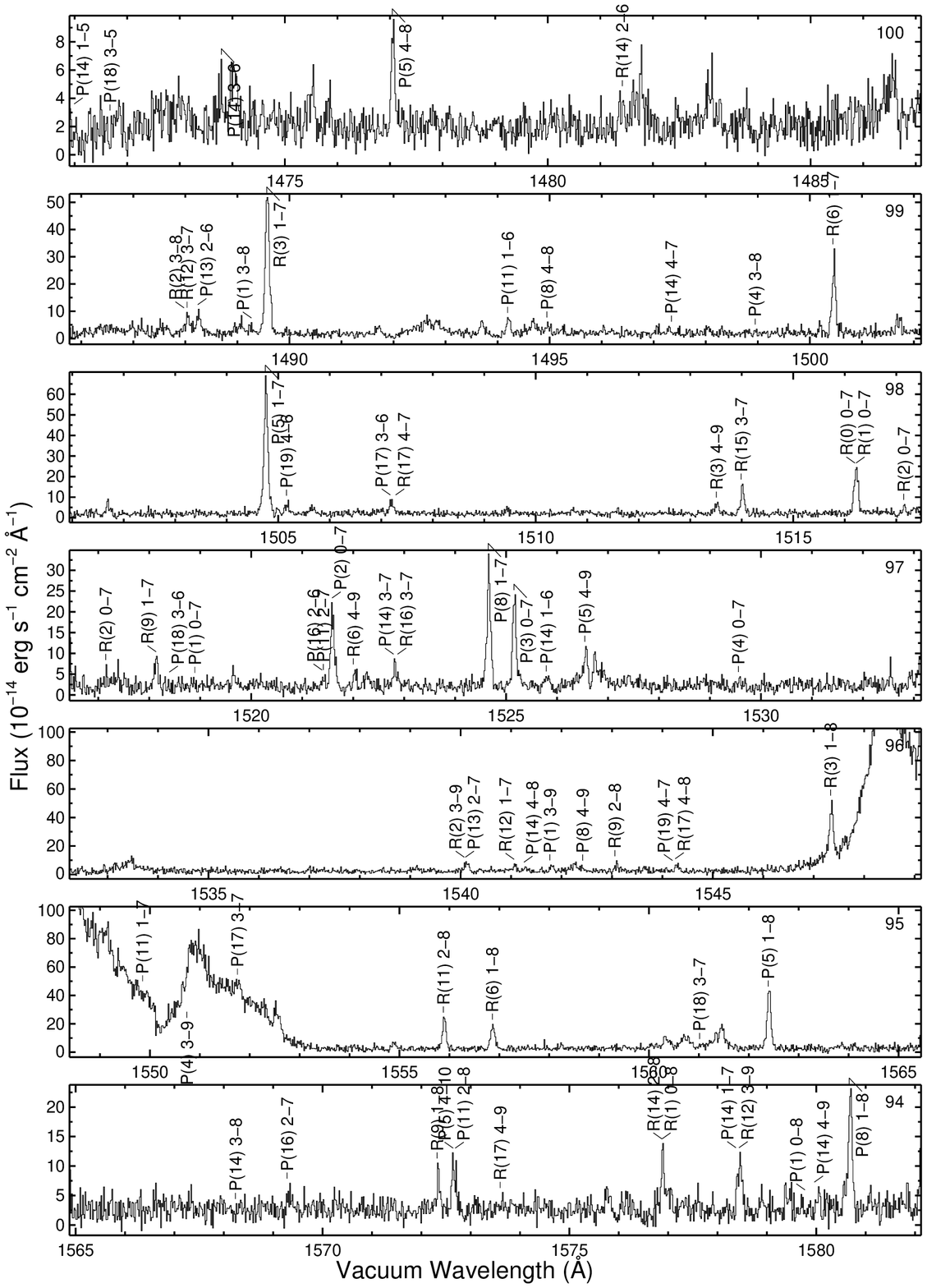}}
\caption{Continued}
\label{fig:herczeg5.eps}
\end{figure}
\setcounter{figure}{0}

\begin{figure}
\resizebox{6.5in}{8.in}{\plotone{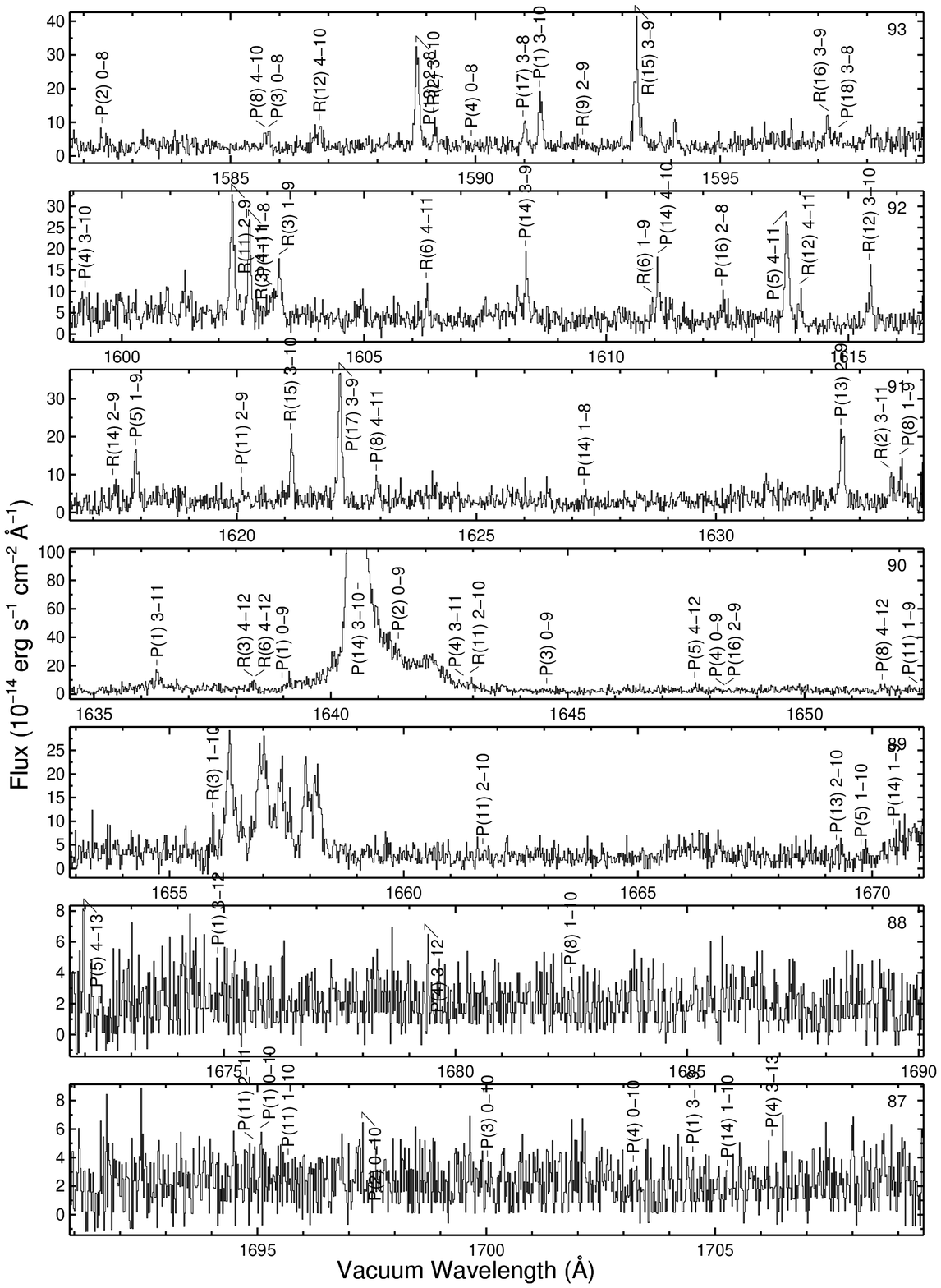}}
\label{fig:herczeg6.eps}
\caption{Continued}
\end{figure}
\setcounter{figure}{1}

\begin{figure}
\resizebox{6.5in}{8.in}{\plotone{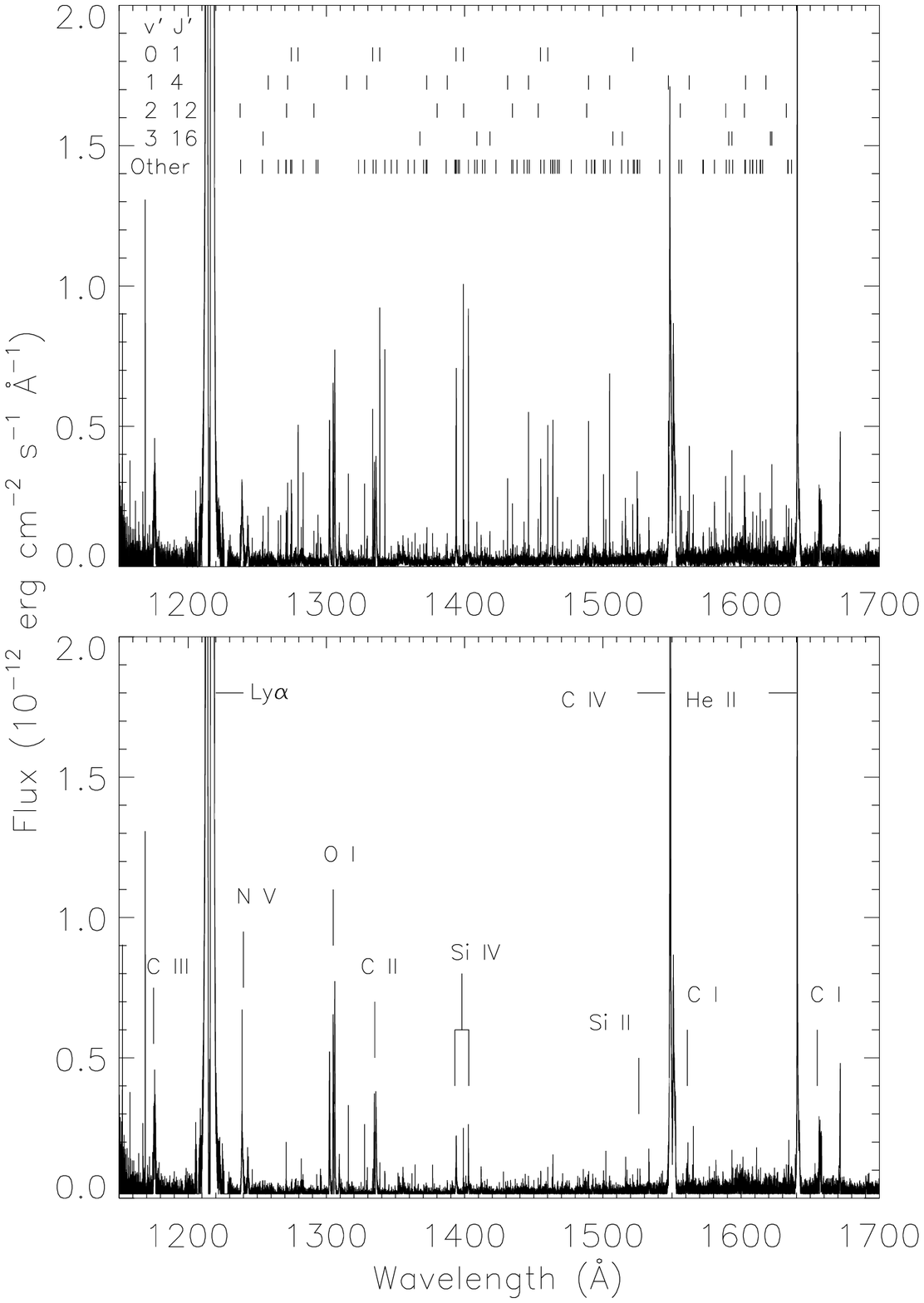}}
\label{fig:ploth2.eps}
\caption{E140M spectrum of TW Hya.  H$_2$ lines are marked (top) from four
upper states, while 
H$_2$ lines in all other
progressions are marked together.  The spectrum without H$_2$ lines is
shown below.}
\end{figure}

\begin{figure}
\plotone{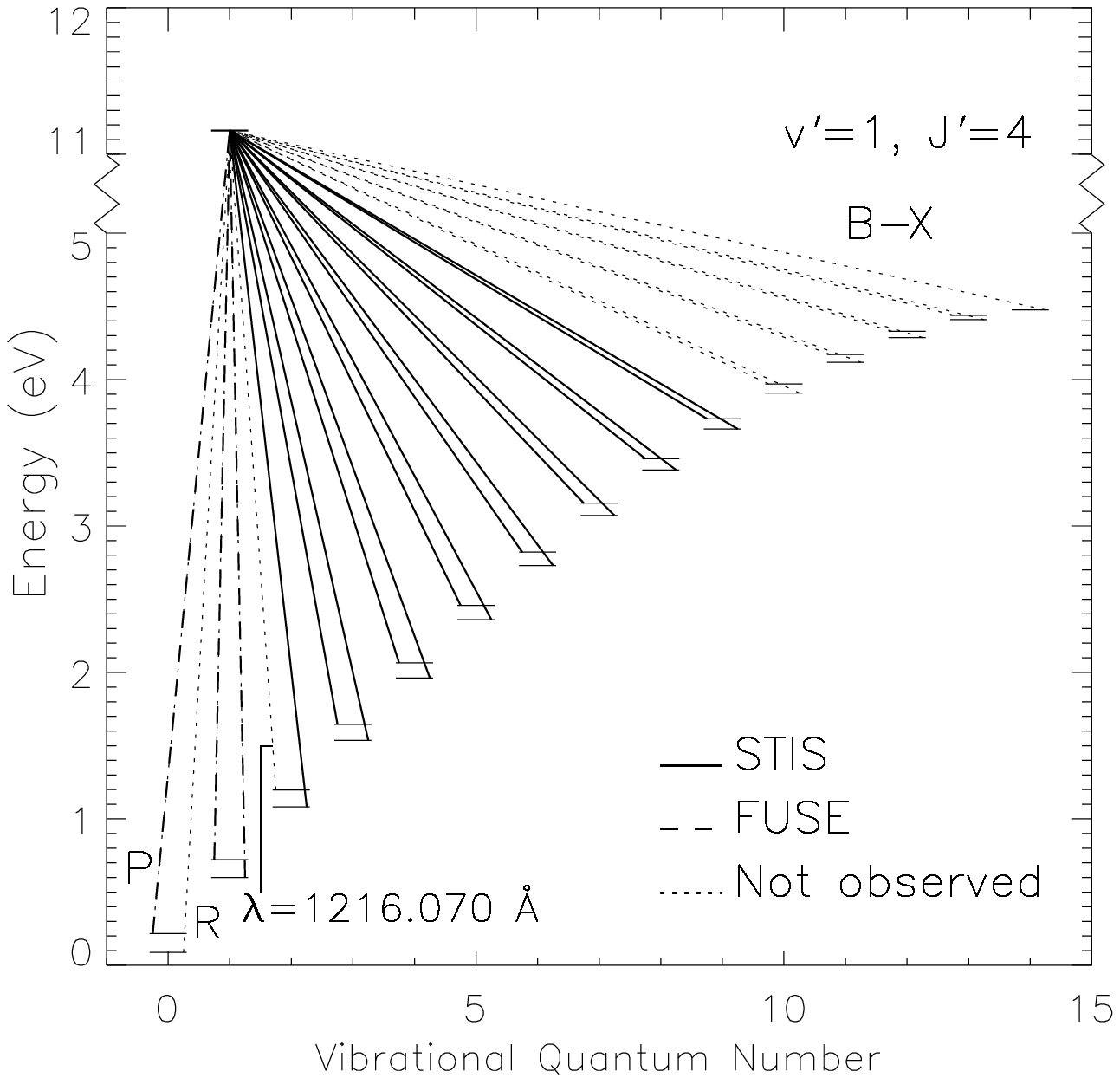}
\label{fig:cascade_ev.eps}
\caption{H$_2$ is pumped by \Lya\, in this example via the 1-2 P(5)
transition at 1216.070 \AA\ to $\Vup=1, \Jup=4$ at $\Ehi=11.4$ eV. From this excited
electronic level, 
H$_2$ returns to many vibrational levels of the 
ground electronic state via $R$
and $P$ transitions.  We detect emission with STIS (solid lines) and \textit{FUSE} 
(dashed lines), but no emission for the transitions indicated by dotted
lines.}
\end{figure}

\begin{figure}
\label{fig:fuseh2.eps}
\plotone{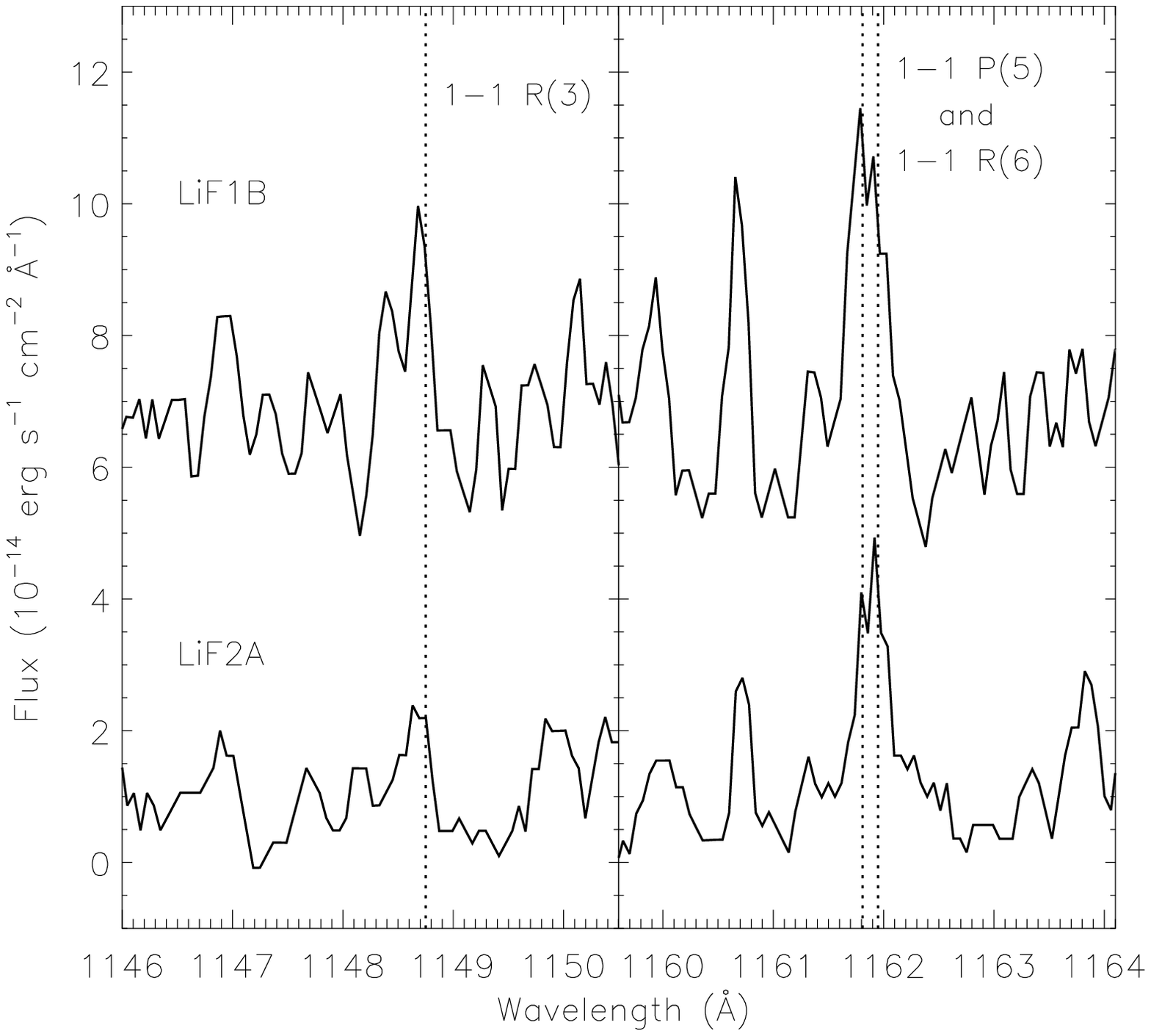}
\caption{\FUSE\ observations of two H$_2$ lines in the LiF1B and LiF2A
channels.  The flux shown here is smoothed over 3 resolution elements.  
The line near 1161.9 \AA\ 
is a blend of two H$_2$ transitions, but its dominated by the 1-1 P(5)
transition.}
\end{figure}

\begin{figure}
\label{fig:twboo.eps}
\resizebox{6.5in}{8.in}{\plotone{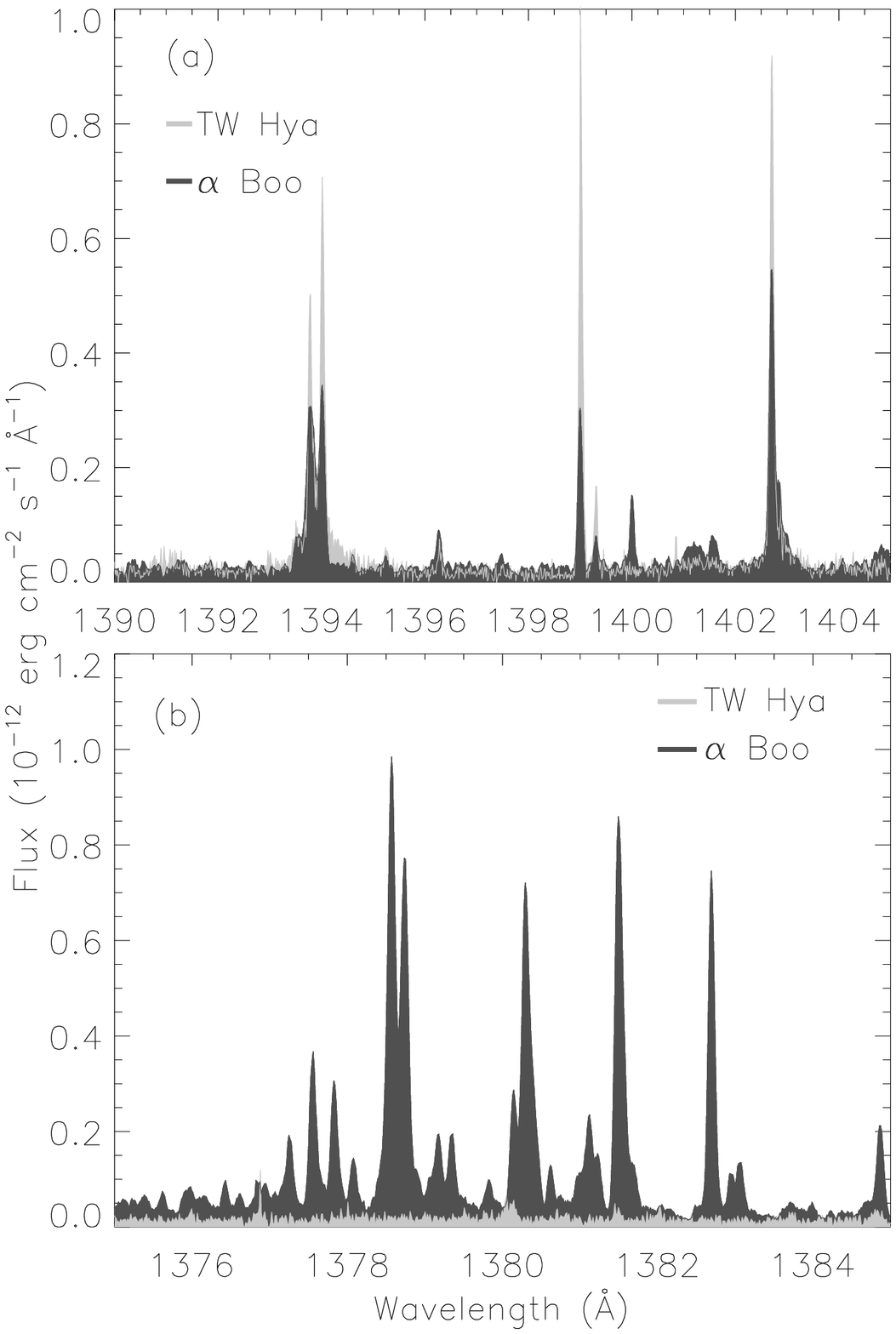}}
\caption{A comparison of \textit{\HST}/STIS E140M spectra of TW Hya and
the evolved giant $\alpha$ Boo \citep{Ayr01}.  Both stars show narrow
H$_2$ features (top) at wavelengths normally dominated by \ion{Si}{4}
emission.  The spectrum of $\alpha$ Boo has many strong CO A-X lines near 
1380 \AA, which are pumped by  the \ion{O}{1} 1304 \AA\ triplet, but TW Hya does not
show these CO lines.}
\end{figure}

\clearpage

\begin{figure}
\resizebox{6.in}{7.5in}{\plotone{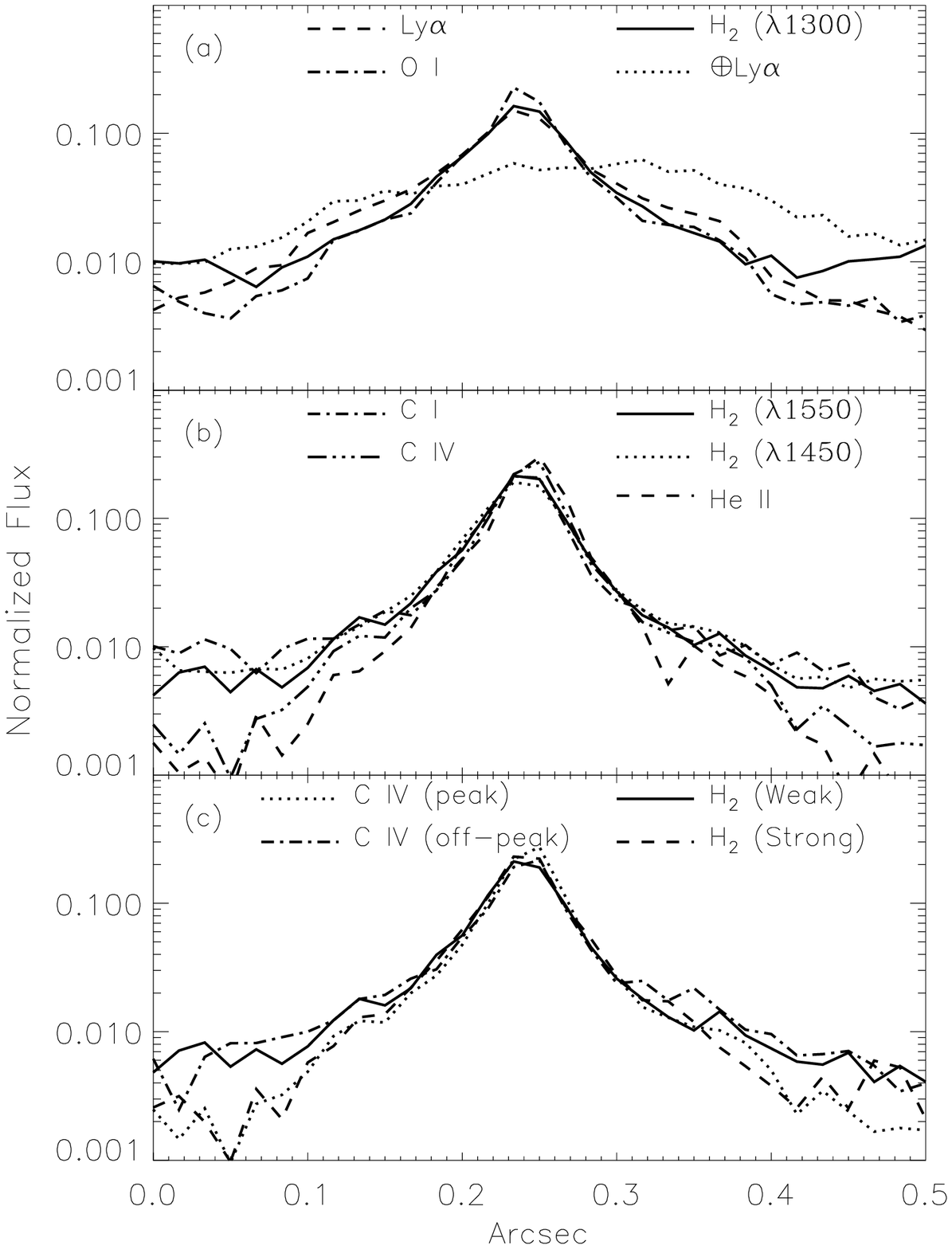}}
\label{fig:spatial_1.eps}
\caption{(a) The spatial extent of H$_2$ emission (solid) for lines below 1350 \AA,
compared with Ly$\alpha$ emission (dashed), geocoronal Ly$\alpha$  emission (dotted), and
\ion{O}{1} emission (dashed-dotted) across the aperture in the
cross-dispersion direction.  (b) The spatial extent of H$_2$ emission (solid) above 1500 \AA,
compared with H$_2$ emission from 1350--1500 \AA\ (dotted), 
  \ion{C}{4} emission (dashed-dotted), \ion{He}{2} emission (dashed) and
\ion{C}{1} emission (dashed-dotted), across the aperture in the
cross-dispersion direction.  (c)  The spatial extent of strong and weak
H$_2$ lines above 1500 \AA\, compared with \ion{C}{4} emission from its
peak and wing.}
\end{figure}

\begin{figure}
\label{fig:emlines.eps}
\resizebox{6.5in}{8.in}{\plotone{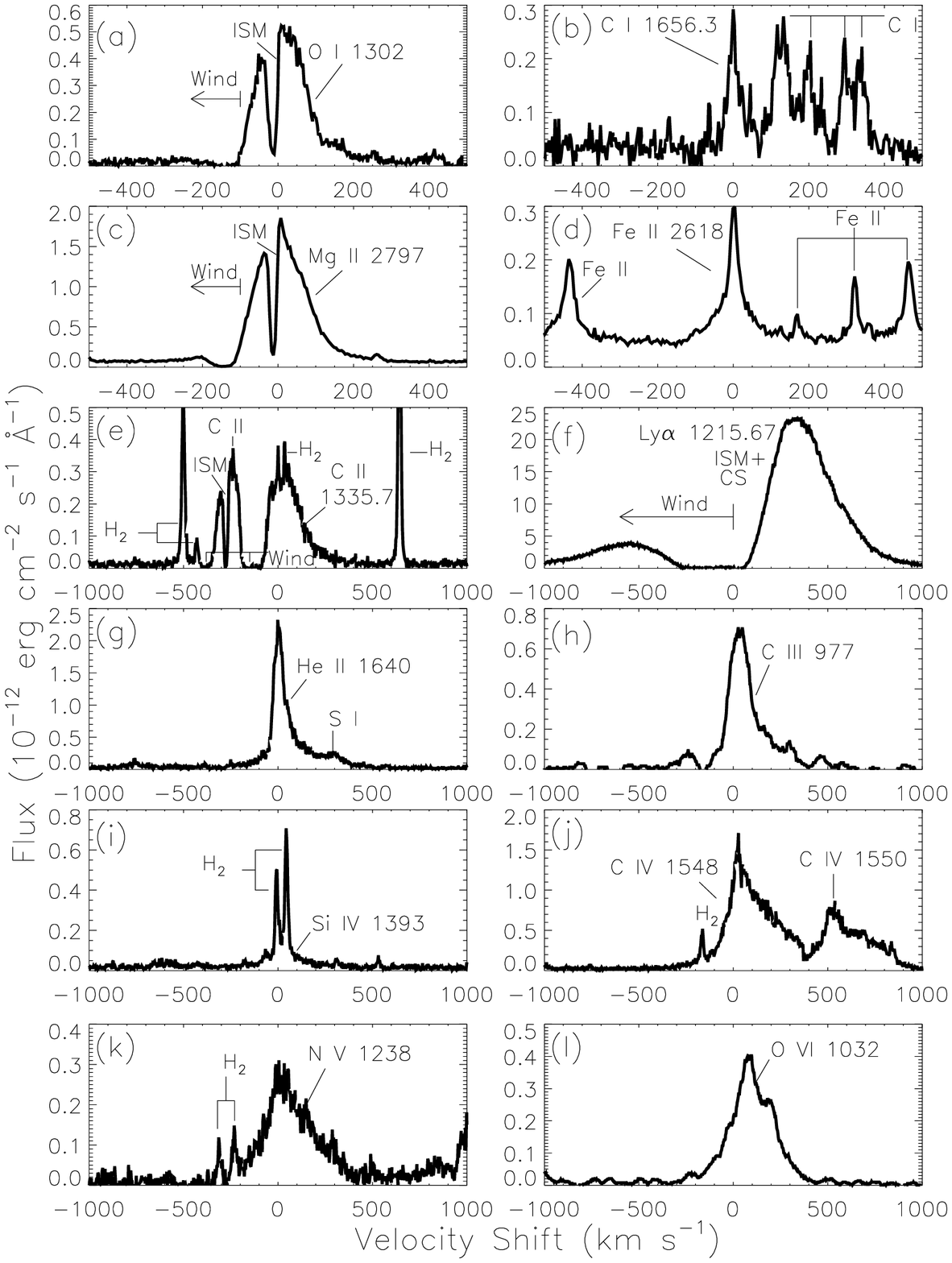}}
\caption{Atomic and molecular lines in the STIS and \textit{FUSE} spectra of TW Hya.}
\end{figure}

\begin{figure}
\plotone{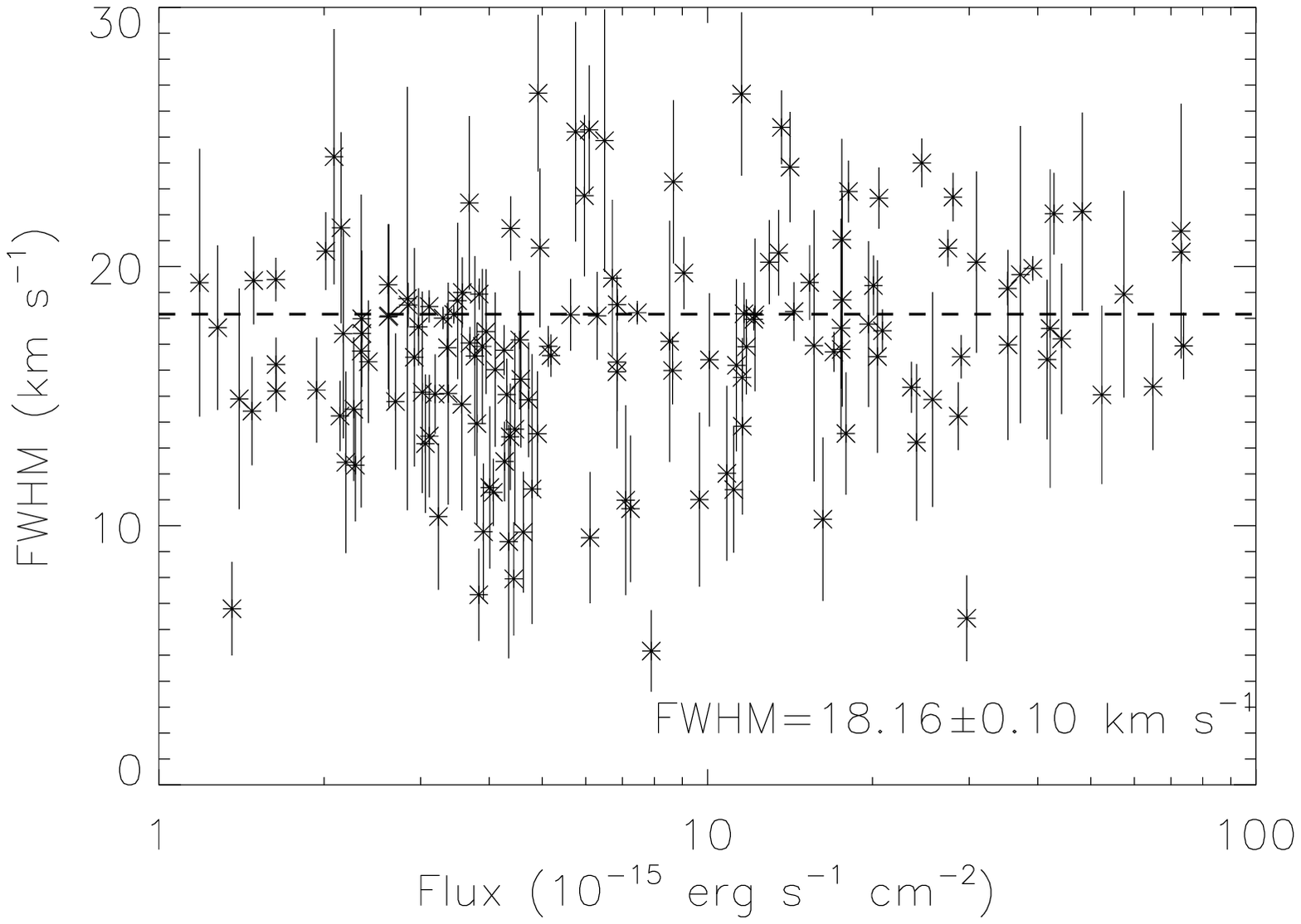}
\label{fig:width.eps}
\caption{The FWHM of the H$_2$ lines, due to instrumental, thermal and rotational
broadening, plotted here as a function of
line flux.  The dashed line is the weighted mean of the H$_2$ FWHM.}
\end{figure}

\begin{figure}
\plotone{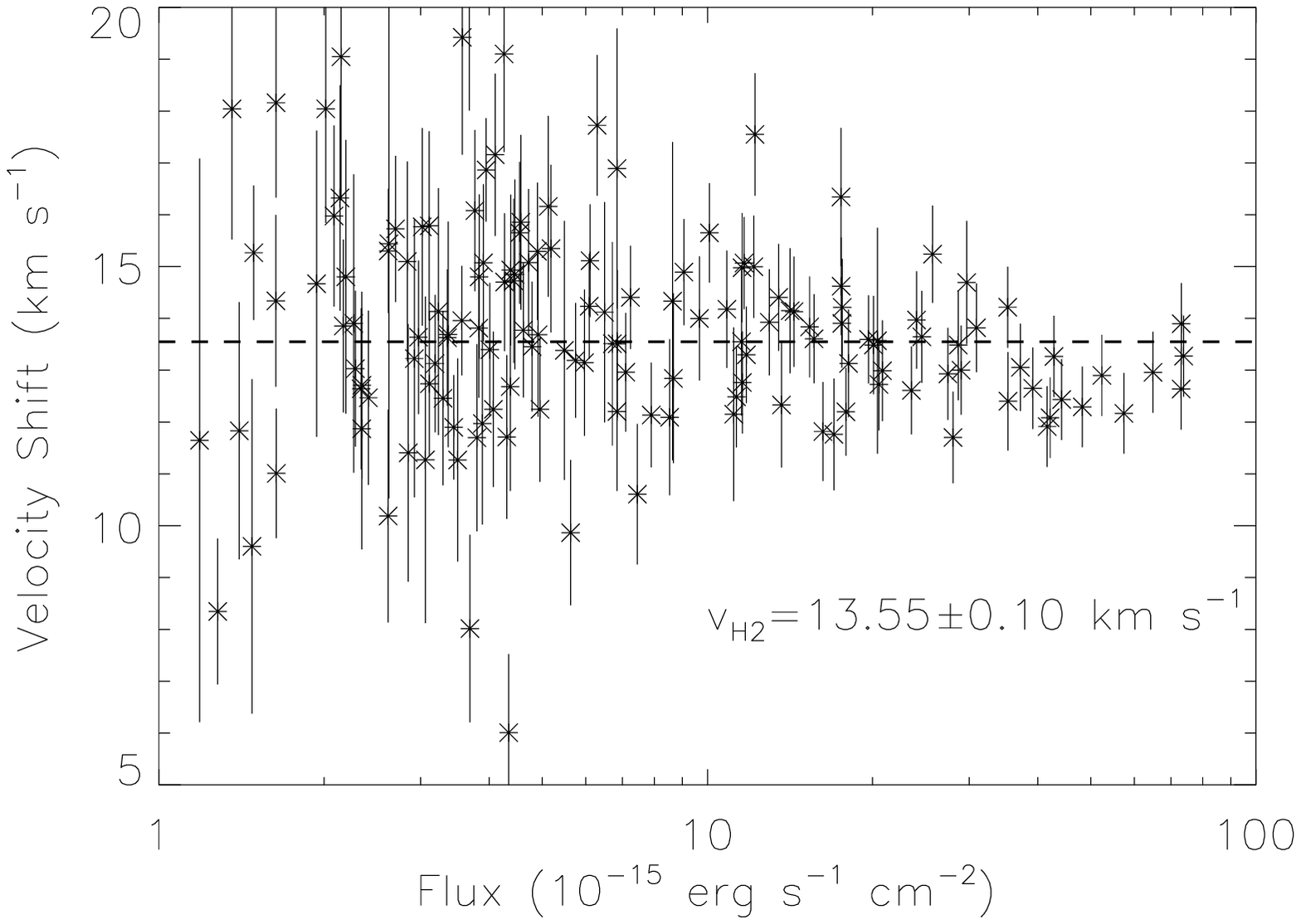}
\label{fig:redshift.eps}
\caption{The velocity shift of H$_2$ lines, plotted here as a function of
line flux, is consistent with the 12-13
\kms\ photospheric radial velocity of TW Hya.  The dashed line is the
weighted mean of the H$_2$ radial velocities.}
\end{figure}

\clearpage

\begin{figure}
\resizebox{5.5in}{7.in}{\plotone{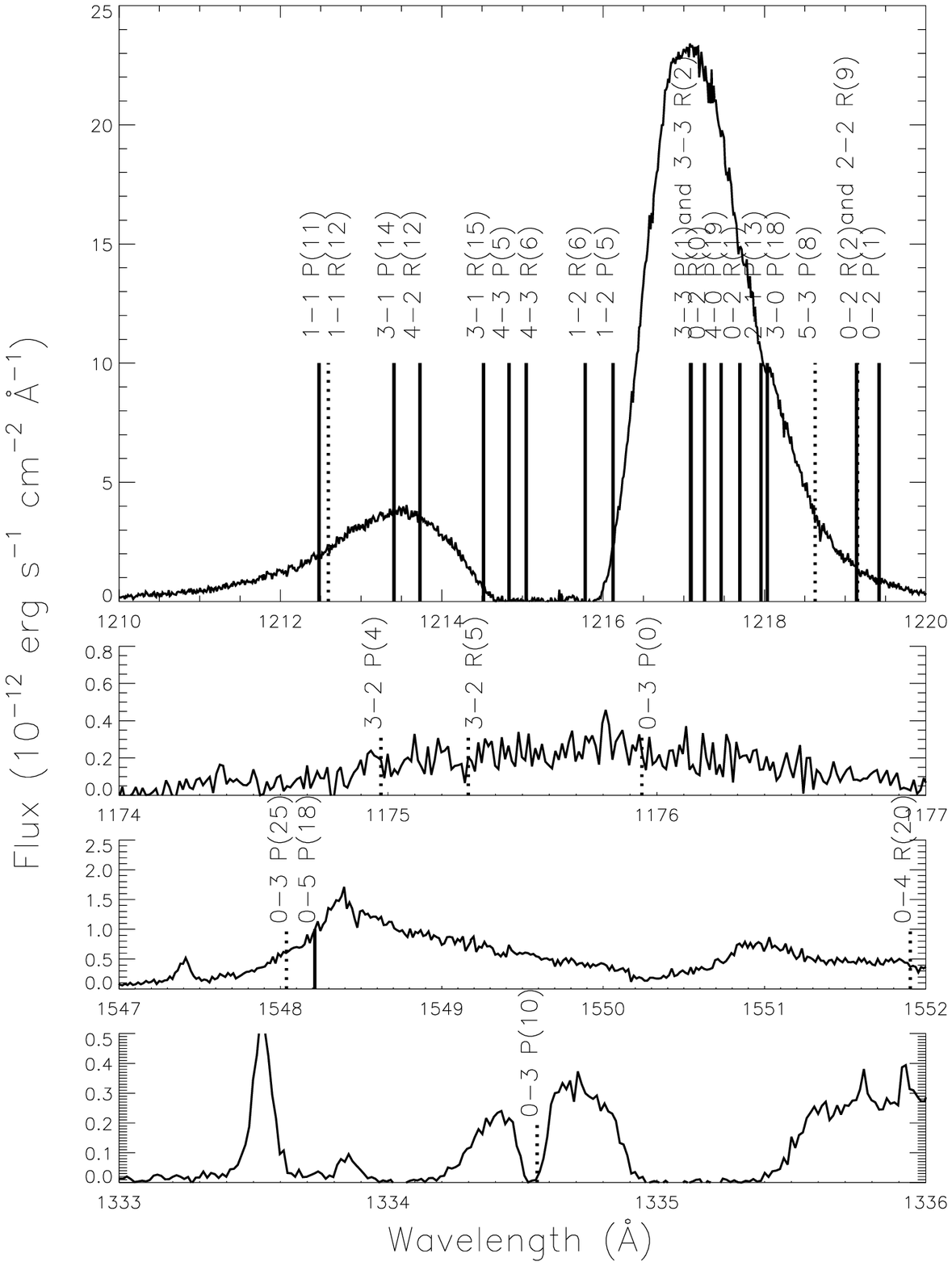}}
\label{fig:pumping1.ps}
\caption{H$_2$ transitions are pumped primarily by \Lya\ (top), although
the \ion{C}{3} 1175 \AA\ multiplet (upper middle), the \ion{C}{4} 1549 \AA\
doublet (lower middle), and the \ion{C}{2} 1334.5 \AA\ line (bottom) also
pump some transitions.  The solid lines indicate
pumping transitions with definite detection of multiple fluorescent lines,
 and the dotted lines indicate transitions where the fluorescent lines are
tentatively detected.}
\end{figure}

\begin{figure}
\plotone{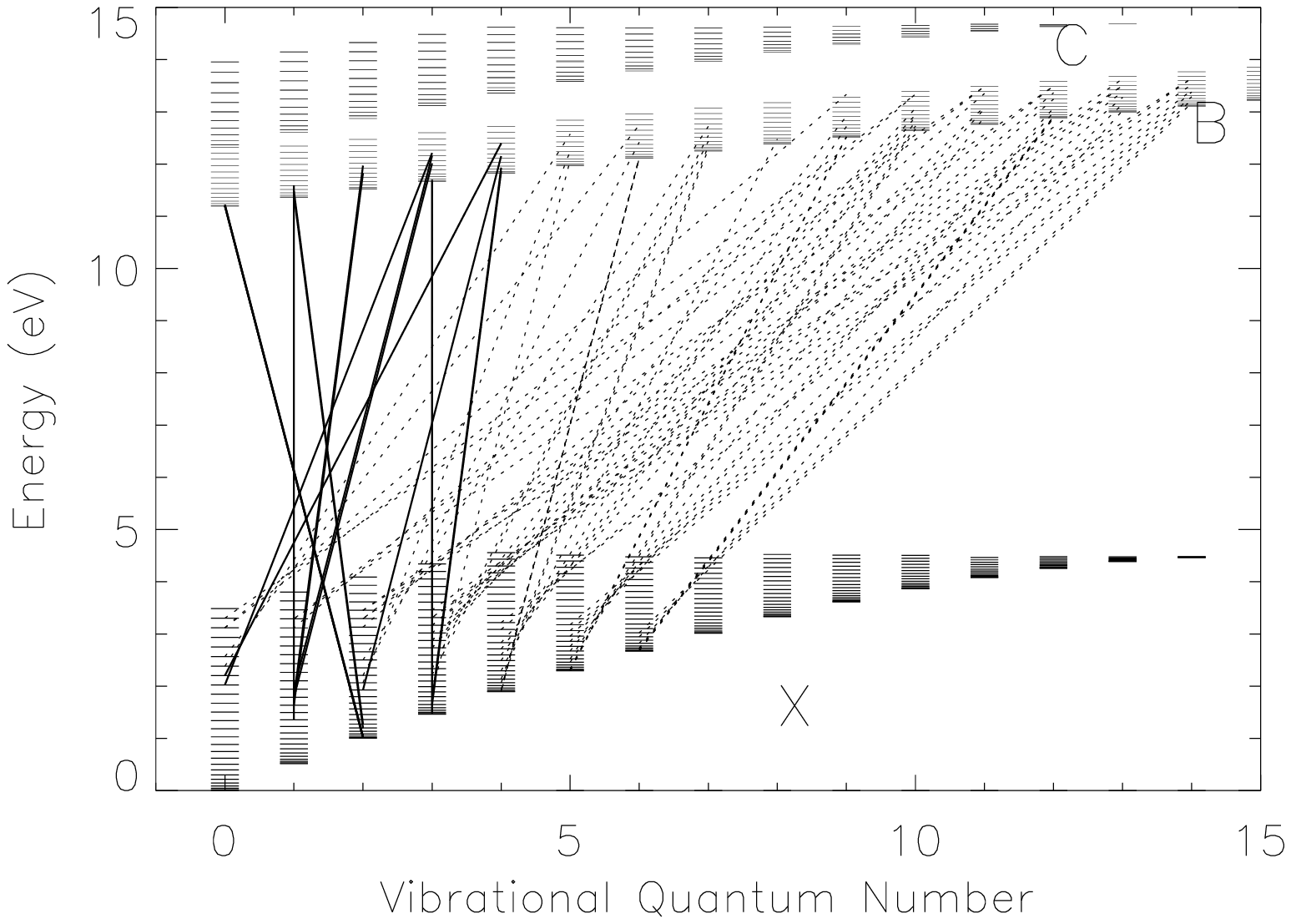}
\label{fig:enlevs_ev.eps}
\caption{Energy levels calculated from transition data in \citet{Abg93} of the ground (X), and excited (B and C) electronic
states.  Lyman-band (B-X) transitions are discussed in this paper, 
while Werner-band (C-X) transitions typically occur in the 
\textit{FUSE} bandpass and have no transitions from low energy levels of
the ground state that are
coincident with \Lya.  We observe fluorescence from upper levels
pumped by Ly$\alpha$ via transitions indicated by the solid lines.  Dashed
lines indicate transitions coincident with Ly$\alpha$ from which
fluorescence is not detected.  We do not detect fluorescence in progressions
pumped from very high energy levels of the ground state, which indicates
that the populations of H$_2$ in the ground electronic state are probably thermal.}
\end{figure}

\clearpage

\begin{table}[t]
\caption{UV observations of TW Hya}
\begin{tabular}{ccccccccc}
Date & Dataset & Grating & Time & Aperture & $\lambda_{\rm min}$ & 
$\lambda_{\rm max}$ & Resolution\\
\hline 
7 May 2000 & O59D01020 &E230M & 1675 s & $0\farcs2 \times 0\farcs2$
 & 2157 \AA& 2965\AA\ & 30000\\
8 May 2000 & O59D01030 &E140M & 2300 s & $0\farcs5 \times 0\farcs5$
 & 1140 \AA& 1735\AA
& 45800\\
3 June 2000 &P1860101000 &\FUSE\ & 2031 s & $30\arcsec \times 30\arcsec$
 & 900 \AA& 1188\AA\ & 15000\\
\hline
\end{tabular}
\label{tab:obs.tab}
\end{table}

\begin{table}
\rotatebox{90}{
\begin{tabular}{ccccccccccc}
\multicolumn{11}{c}{Table 2: H$_2$ Progressions$^1$}\\
\multicolumn{3}{c}{\underline{Pumped by 0-2 P(1) 1219.368 \AA}} &&  \multicolumn{3}{c}{\underline{Pumped by 0-2 R(0) 1217.205 \AA}}  && \multicolumn{3}{c}{\underline{Pumped by 0-2 R(1) 1217.643 \AA}}\\
1335.921 & 0-4 P(1) &    3.4(0.8)   &&  1274.589 & 0-3 R(0) &   27.4(1.1)  && 1274.980 & 0-3 R(1) &   24.6(1.0)\\
1396.281 & 0-5 P(1) &    3.3(0.5)   &&  1279.518 & 0-3 P(2) &   39.2(1.1)  && 1283.161 & 0-3 P(3) &   28.0(1.2)\\
1457.474 & 0-6 P(1) &    3.7(0.5)   &&  1333.533 & 0-4 R(0) &   42.8(1.2)  && 1333.851 & 0-4 R(1) & 7.9(0.7)\\
&& &&  1338.630 & 0-4 P(2) &   73.1(1.3)  && 1342.314 & 0-4 P(3) &   64.9(1.3)\\
 \multicolumn{3}{c}{\underline{Pumped by 0-2 R(2) 1219.089 \AA}}   &&  1393.785 & 1-5 R(0) &   35.3(1.7)  && 1394.021 & 0-5 R(1) &   52.4(1.9)\\
1346.970 & 0-4 P(4) &    3.8(0.5)    &&  1399.013 & 1-5 P(2) &   73.8(1.7)  && 1402.707 & 1-5 P(3) &   73.1(1.8)\\
1395.255 & 0-5 R(2) &    2.4(0.4)    &&  1454.892 & 1-6 R(0) &   20.8(1.1)  && 1455.038 & 1-6 R(1) &   30.9(1.4)\\
1407.350 & 0-5 P(4) &    3.0(0.5)    &&  1460.223 & 2-6 P(2) &   41.6(1.5)  && 1463.885 & 1-6 P(3) &   42.1(1.4)\\
1468.454 & 0-6 P(4) &    3.4(0.6)    &&  1516.271 & 0-7 R(0) &   21.2(b)  && 1516.271 & 0-7 R(1) &   21.2(b)\\
&&   &&  1521.647 & 2-7 P(2) &   16.2(1.1)  && 1525.215 & 2-7 P(3) &   17.9(1.0)\\
 \multicolumn{3}{c}{\underline{Pumped by 0-5 P(18) 1548.146 \AA}}   &&  &&    && \\
1437.868 & 0-3 P(18) &    1.6(0.5)   &&   \multicolumn{3}{c}{\underline{Pumped by 1-2 P(5) 1216.070 \AA}}  &&  \multicolumn{3}{c}{\underline{Pumped by 1-2 R(6) 1215.726 \AA}}\\
1446.791 & 0-4 R(16) &    4.4(0.5)   &&  1148.701$^2$ & 0-1 R(3) &    4.6(0.9)    && 1237.918 & 0-2 P(8) &   11.5(1.3)  \\
1493.748 & 0-4 P(18) &    4.7(0.7)   &&  1161.875$^2$ & 0-1 P(5) &   10.9(b)    && 1271.074 & 0-3 R(6) &   14.1(1.0)  \\
1501.748 & 0-5 R(16) &    3.8(0.6)   &&  1202.449 & 0-2 R(3) &   11.3(2.2)   && 1293.927 & 0-3 P(8) &   13.0(0.8)  \\
1554.948 & 0-6 R(16) &    4.3(0.8)   &&  1257.883 & 0-3 R(3) &   18.1(0.9)   && 1327.623 & 0-4 R(6) &    6.1(0.7)  \\
&&   &&  1271.979 & 1-3 P(5) &   20.5(1.0)   && 1351.100 & 0-4 P(8) &    2.8(0.4)  \\
 \multicolumn{3}{c}{\underline{Pumped by 1-1 P(11) 1212.425 \AA}}   &&  1314.690 & 1-4 R(3) &   12.2(0.7)   && 1409.024 & 1-5 P(8) &    2.2(0.5)  \\
1292.507 & 0-3 R(9) &    4.9(0.5)    &&  1329.184 & 1-4 P(5) &    7.5(0.7)   && 1442.920 & 1-6 R(6) &   11.3(0.7)  \\
1323.303 & 0-3 P(11) &    3.8(0.6)   &&  1372.550 & 2-5 R(3) &    3.2(0.4)   && 1467.147 & 1-6 P(8) &   17.6(1.1)  \\
1380.131 & 1-4 P(11) & 3.4(b)   &&  1387.422 & 2-5 P(5) &    7.1(0.6)   && 1500.511 & 2-7 R(6) &   19.7(1.2)  \\
1462.201 & 0-6 R(9) &    3.8(0.6)    &&  1431.072 & 3-6 R(3) &   29.0(1.2)   && 1524.712 & 2-7 P(8) &   23.5(1.3)  \\
1494.260 & 0-6 P(11) &    4.8(0.6)   &&  1446.178 & 3-6 P(5) &   44.2(1.5)   && 1556.921 & 3-8 R(6) &   17.0(1.1)  \\
1518.184 & 0-7 R(9) &    5.6(0.9)    &&  1489.625 & 3-7 R(3) &   48.2(1.7)   && 1580.743 & 3-8 P(8) &   17.5(1.8)  \\
1572.402 & 1-8 R(9) &    4.5(0.9)    &&  1504.812 & 3-7 P(5) &   57.5(2.3)   && 1633.905 & 3-9 P(8) &    5.5(1.5)  \\
1603.314 & 1-8 P(11) &    9.7(2.2)   &&  1547.398 & 3-8 R(3) &   35.3(2.7)   && \\
&&   &&  1562.457 & 4-8 P(5) &   37.2(1.7)   && \multicolumn{3}{c}{\underline{Pumped by 4-3 R(6) 1214.995 \AA}}\\ 
 \multicolumn{3}{c}{\underline{Pumped by 3-0 P(18) 1217.982 \AA}}   &&  1603.164 & 4-9 R(3) &   11.2(2.2)   && 1393.003 & 0-6 P(8) &    1.2(0.3) \\
1268.890 & 0-1 P(18) &    1.4(0.4)    &&  1617.952 & 4-9 P(5) &   11.6(1.9)   && 1522.086 & 0-9 R(6) &    2.1(0.5) \\
1597.239 & 0-9 R(16) &    4.3(1.0)    &&    &&    && 1606.359 & 0-11 R(6) &    4.4(0.9) \\
1597.514 & 0-8 P(18) &    2.0(0.9)    &&    && \\
\end{tabular}}
\end{table}

\begin{table}
\rotatebox{90}{
\begin{tabular}{ccccccccccc}
\multicolumn{11}{c}{Table 2 (continued):}\\
\multicolumn{3}{c}{\underline{Pumped by 2-1 P(13) 1217.904 \AA}}  &&
\multicolumn{3}{c}{\underline{Pumped by 2-1 R(14) 1218.521 \AA}}   &&
\multicolumn{3}{c}{\underline{Pumped by 3-3 P(1) 1217.038 \AA}}\\
1237.589 & 0-2 R(11) &    8.7(1.1)  &&  1257.460 & 0-1 P(16) &    2.7(0.5)    &&1270.631 & 0-4 P(1) &    3.1(0.5) \\
1271.238 & 0-2 P(13) &   13.5(0.9)   &&  1310.622 & 0-2 P(16) &    4.1(0.6)   &&1380.131 & 3-6 P(1) & 3.4(b)\\
1290.961 & 0-3 R(11) &    9.1(0.6)   &&  1323.741 & 0-3 R(14) &    2.1(0.4)   &&1435.112 & 0-7 P(1) &    4.0(0.5) \\
1380.042 & 0-4 P(13) &    2.3(0.5)   &&  1470.513 & 0-5 P(16) &    1.9(0.5)   &&1591.385 & 3-10 P(1) & 14.4(1.5)\\
1399.304 & 0-5 R(11) &   12.1(0.9)   &&  1569.340 & 0-7 P(16) &    3.1(0.9)   &&1636.401 & 0-11 P(1) &    7.8(1.8) \\
1434.598 & 1-5 P(13) &   15.6(0.9)   &&  1612.458 & 1-8 P(16) &    3.6(0.9)   &&\\
1453.160 & 1-6 R(11) &   15.4(0.9)   &&  1617.505 & 1-9 R(14) &    4.1(1.3)   &&\multicolumn{3}{c}{Pumped by 3-3 R(2) 1217.031 \AA}\\
1488.306 & 1-6 P(13) &    6.8(0.8)   &&    &&    && \multicolumn{3}{c}{\underline{and 3-2 P(4) 1174.923 \AA}}\\
1540.174 & 2-7 P(13) & 5.2(b)  &&   \multicolumn{3}{c}{\underline{Pumped by 3-1 P(14) 1213.356 \AA}}  &&1434.143 & 0-7 R(2) &    1.5(0.4)\\
1555.950 & 2-8 R(11) &   20.1(1.2)   &&   1370.343 & 0-4 P(14) &    4.0(0.5)  &&1445.258 & 0-7 P(4) &    3.1(0.5)\\
1588.867 & 2-8 P(13) &   24.1(1.6)   &&  1386.577 & 0-5 R(12) &    3.4(0.5)  &&1540.174 & 3-9 R(2) & 5.2(b)\\
1602.336 & 3-9 R(11) &   28.7(2.5)   &&  1422.670 & 0-5 P(14) &    1.6(0.4)  &&1589.226 & 3-10 R(2) & 4.6(0.9)\\
1632.688 & 3-9 P(13) &   17.5(2.0)  &&  1488.100 & 0-7 R(12) &    6.7(1.0)   &&1633.710 & 0-11 R(2) &    2.9(0.8) \\
  &&    &&  1522.875 & 1-7 P(14) &    5.1(0.8)   &&\\
\multicolumn{3}{c}{\underline{Pumped by 3-1 R(15) 1214.465 \AA}}  &&  1578.435 & 3-9 R(12) & 8.9(b)  && \multicolumn{3}{c}{\underline{Pumped by 4-2 R(12) 1213.677 \AA}}\\
1254.188 & 0-1 P(17) &   11.7(0.8)   &&  1608.409 & 1-9 P(14) &    8.5(1.1)   &&1349.699 & 0-4 P(14) &    0.8(0.2)\\
1265.232 & 0-2 R(15) &   13.6(0.9)   &&  1615.507 & 3-10 R(12) & 6.8(1.4)  &&1415.400 & 0-6 R(12) &    1.5(0.3)\\
1367.681 & 0-4 R(15) &    4.3(0.6)   &&    &&    &&1509.491 & 0-8 R(12) &    1.3(2.9)\\
1408.876 & 0-4 P(17) &   11.6(0.7)   &&   \multicolumn{3}{c}{\underline{Pumped by 4-3 P(5) 1214.781 \AA}}  &&1541.331 & 0-8 P(14) &    2.2(0.8)\\
1418.291 & 0-5 R(15) &   10.9(0.7)   &&  1253.716 & 0-4 R(3) &    6.0(0.7)   &&1611.132 & 4-10 P(14) & 8.6(1.2)\\
1507.250 & 1-6 P(17) &    4.4(0.7)   &&  1266.930 & 4-4 P(5) & 12.9(b)  &&1614.063 & 1-11 R(12) &    4.5(1.1) \\
1514.067 & 1-7 R(15) &   10.1(0.9)   &&  1359.146 & 1-6 R(3) &    6.5(0.8)   &&\\
1591.064 & 1-8 P(17) &    6.3(1.2)   &&  1372.768 & 1-6 P(5) &    7.2(0.7)   &&  \multicolumn{3}{c}{\underline{Pumped by 4-0 P(19) 1217.410 \AA}}\\
1593.339 & 2-9 R(15) &   25.7(1.9)   &&  1463.641 & 1-8 R(3) &    2.3(0.6)   &&1266.930 & 4-1 P(19) & 12.9(b)\\
1621.200 & 3-10 R(15) & 11.8(1.5)    &&  1477.106 & 2-8 P(5) &    5.8(0.7)   &&1274.085 & 0-2 R(17) &    5.0(0.6)  \\
1622.205 & 3-9 P(17) &   29.7(2.3)   &&  1513.566 & 2-9 R(3) &    4.3(0.7)  &&1372.131 & 0-4 R(17) &    4.9(0.6)  \\
&&
&&  1526.622 & 3-9 P(5) &    6.1(1.0)   &&1414.761 & 0-4 P(19) &
4.6(0.6)  \\
\multicolumn{3}{l}{$^1$Observed $\lambda$, line ID, \& Flux($\sigma$)}&&  1572.709 & 4-10 P(5) & 6.8(1.1)  && 1465.261 & 0-6 R(17) &    5.2(0.8)  \\
\multicolumn{3}{l}{Flux, $\sigma$ in 10$^{-15}$ erg cm$^{-2}$ s$^{-1}$}
  &&  1602.689 & 3-11 R(3) &   17.6(2.0)   &&1505.222 & 0-6 P(19) &    3.9(0.8)  \\
\multicolumn{3}{l}{$^2$Observed with \FUSE} 
 &&  1613.784 & 3-11 P(5) &   20.4(1.5)   &&1544.345 & 1-8 R(17) &    3.6(0.8)  \\
\multicolumn{3}{l}{(b) indicates blend}  \\
\end{tabular}}
\end{table}

\setcounter{table}{2}

\begin{center}

\begin{table}
\caption{Downward transitions from $\Vup=0, \Jup=17$ arranged by $A_{ul}$}
\begin{tabular}{cclllccclll}
\hline
$\lambda$ & A$_{ul}$ & $\Vlo$ & $\Jlo$ & Flux$^1$
&&$\lambda$ & A$_{ul}$ & $\Vlo$ & $\Jlo$ & Flux$^1$\\
\AA\ & (s$^{-1}$) & & & $10^{-15}$  &&\AA\ & (s$^{-1}$) & & & $10^{-15}$ \\
\hline
\hline
1501.674&   2.178e+08&  5&16&5.01 & & 1604.905&   5.346e+07&  7&16& $<5$  \\
1493.673&   2.107e+08&  4&18&4.63   && 1335.299&   4.231e+07&  2&16& $<1.5$\\
1446.719&   1.948e+08&  4&16&4.35  & & 1647.325&   1.914e+07&  7&18& $<2.5$ \\
1548.146&   1.909e+08&  5&18& $<20$ & &  1325.281&   1.501e+07&  1&18& $<2.5$  \\
1554.849&   1.497e+08&  6&16&5.42  & & 1280.162&   9.346e+06&  1&16 & $<2$ \\
1437.781&   1.435e+08&  3&18&3.03   &&  1650.012&   5.812e+06&  8&16& $<5$ \\
1391.008&   1.132e+08&  3&16& $<3$  & & 1269.897&   1.648e+06&  0&18& $<2$  \\
1599.932&   9.589e+07&  6&18& $<5$  &  & 1226.004&   9.338e+05&  0&16& $<8$\\
1381.413&   6.097e+07&  2&18& $<3$   & &  1687.919&   2.402e+05&  8&18& $<3$ \\
\hline
\multicolumn{5}{l}{$^1$ erg s$^{-1}$ cm$^{-2}$}\\
\end{tabular}
\label{tab:vu0ju17.tab}
\end{table}

\begin{table}
\caption{Tentatively identified H$_2$ lines}
\label{tab:tentative}
\begin{tabular}{cccccc}
\hline
$\lambda_{\rm obs}$ & ID & Flux$^1$ & Pump ID & Pump $\lambda_{\rm calc}$ &
\Elo\ (eV)\\
\hline
\hline
    1412.887 &      2-5 P(11)  &   2.6(0.6) & 2-2 R(9) & 1219.101 & 1.56\\
    1465.110 &      0-6 R(5)  &   2.1(0.6) & 0-1 R(5) & 1173.981$^2$ & 0.72\\
    1363.575 &      0-4 R(8)  &   2.3(0.4) & 0-3 P(10) & 1334.497 & 2.10\\
    1393.506 &     0-4 P(10)  &   2.8(1.0) & 0-3 P(10) & 1334.497 & 2.10\\
    1404.040 &     0-4 P(11)  &   2.3(0.5) & 0-0 P(11) & 1175.893 & 0.88\\
    1522.331 &     0-6 P(11)  &   3.7(0.8) & 0-0 P(11) & 1175.893 & 0.88\\
    1500.244 &     0-3 P(22)  &   1.6(0.4) & 0-4 R(20) & 1551.836$^3$ & 3.76\\
    1491.761 &      1-7 P(2)  &   3.5(0.6) & 1-8 P(2) & 1550.085$^2$ & 3.36\\
    1498.097 &     0-2 P(25)  &   1.4(0.5) & 0-3 P(25) & 1547.971 & 4.20\\
    1542.322 &     0-4 R(23)  &   3.0(0.8) & 0-3 P(25) & 1547.971 &  4.20\\
    1594.139 &     0-4 P(25)  &   2.0(0.9) & 0-3 P(25) & 1547.971 &  4.20\\
    1359.052 &     1-3 P(14)  &   2.6(0.7) & 1-1 R(12) & 1212.543 & 1.49\\
    1541.125 &     1-7 R(12)  &   2.3(0.6) & 1-1 R(12) & 1212.543 & 1.49\\
1578.435 & 1-7 P(14) & 8.9$^4$& 1-1 R(12) & 1212.543 & 1.49\\
    1593.826 &     3-10 R(5)  &   3.2(0.8) & 3-2 R(5) & 1175.248 & 1.20\\
    1608.238 &     5-12 R(6)  &   2.6(0.8) & 5-3 P(8) & 1218.575 & 1.89\\
\hline
\multicolumn{5}{l}{$^110^{-15}$ erg cm$^{-2}$ s$^{-1}$}\\
\multicolumn{5}{l}{$^2$Unlikely pumping mechanisms}\\
\multicolumn{5}{l}{$^3$Also 0-5 P(22) at 1550.826 \AA}\\
\multicolumn{5}{l}{$^4$Flux uncertain due to line blend}
\end{tabular}
\end{table}

\clearpage
\begin{table}[t]
\caption{Gaussian-shaped lines in the E140M spectrum of TW Hya}
\begin{tabular}{ccccccccc}
\hline
ID & $\lambda_{obs}$ & $v$ & $\sigma(v)$ & FWHM & $\sigma$(FWHM) & Flux & $\sigma$(F) & $\chi^2$\\
 & \AA\ & \kms & \kms & \kms & \kms
 & \multicolumn{2}{c}{(10$^{-15}$ erg cm$^{-2}$ s$^{-1}$ \AA$^{-1}$)} &\\
\hline
\hline
\ion{C}{3}& 1175.734 & - & - & 417 &  10.3 & 412 & 11 & 1.28\\
\ion{S}{1} & 1295.696 &     9.7 &   1.6 &  47.7 &    3.70 & 14.85 & 0.91
 & 1.05\\
\ion{S}{1} & 1296.231 &     13.9 &   2.1 &  51.4 &    4.63 & 11.89 & 0.83
 & 1.05\\
\ion{Si}{2} & 1309.333 &   13.1   &   1.8 & 133.6 &    3.90 & 63.20 & 1.91
 & 1.55\\
 ? & 1315.725 &  -    &  1.8  &   9.6 &    2.3 &  2.14 & 0.37
 & 1.13\\
\ion{C}{1} & 1328.896 &     14.4 &   1.3 &  10.8 &    2.71 &  2.92 & 0.42
 & 1.19\\
\ion{C}{1} & 1329.183 &   18.7   & 2.1   &  25.7 &    2.48 &  8.33 & 0.62
 & 1.19\\
\ion{C}{1}& 1329.652 &     14.4 &   1.8 &  34.5 &    3.84 &  8.03 & 0.79
 & 1.24\\
\ion{Cl}{1} & 1351.709 & 11.8 & 3.9 & 49.5 & 9.0 & 4.81 & 0.66 & 1.01\\
? & 1355.434 & - & 1.9&17.0 & 1.8 & 6.17 & 0.69 & 0.93\\
? & 1355.642 & - & 2.4 &14.4 & 4.8 & 1.86 & 0.51 & 0.93\\
\ion{O}{1} & 1355.844 & 54.5 & 3.0 & 48.5 & 7.8 & 7.53 & 0.96 & 0.80\\
? & 1356.200 & -  & 3.2 & 15.2 & 5.4 & 1.39 & 0.46 & 0.80\\
? & 1360.228 &   -   &   7.3 &  27.8 &    8.60 &  2.76 & 0.66
 & 0.79\\
\ion{Si}{4}& 1393.935 &    42.8 &   2.8 & 168.3 &    6.46 & 75.92 & 4.51
 & 1.62\\
\ion{Si}{4} & 1402.828 &    9.4 &   3.4 & 130.2 &    8.13 & 34.14 & 2.74
 & 1.16\\
\ion{S}{1} & 1425.084 &  11.4   &   1.7 &  10.9 &    3.58 &  2.07 & 0.46
 & 1.26\\
? & 1436.618 &   -   &   1.2 &  12.7 &    2.92 &  3.51 & 0.57
 & 1.02\\
\ion{C}{1} & 1463.351 &  3.0    &   2.2 &  31.0 &    4.31 &  8.99 & 1.07
 & 0.92\\
\ion{C}{1} & 1560.403 &  18.1    & 1.7   &  28.6 &    4.42 & 12.97 & 1.64
 & 1.00\\
\ion{C}{1} & 1560.784 &  19.6    & 1.9   &  48.8 &    4.42 & 22.01 & 2.03
 & 1.00\\
\ion{C}{1}& 1561.460 &  23.1     &  1.6  &  62.6 &    3.84 & 36.79 & 2.57
 & 1.00\\
\ion{C}{1} & 1656.362 &   17.2   &  1.6  &  52.5 &    3.44 & 53.15 & 4.17
 & 1.00\\
\ion{C}{1} & 1657.029 &  3.8    &  1.8  &  49.1 &    3.98 & 62.29 & 4.36
 & 1.28\\
\ion{C}{1} & 1657.448 &   12.5   & 2.9   &  60.5 &    7.60 & 43.67 & 3.90
 & 1.28\\
\ion{C}{1} & 1657.967 & 10.9    &  1.6  &  25.3 &    3.62 & 27.71 & 3.05
 & 1.28\\
\ion{C}{1}& 1658.204 & 15.0    & 2.0  &  36.0 &    4.88 & 32.75 & 3.27
 & 1.28\\
\hline
\end{tabular}
\label{tab:linelist.tab}
\end{table}

\begin{table}[h]
\label{tab:pumplist.tab}
\caption{H$_2$ pumping transitions}
\begin{tabular}{lcccccccc}
\hline
\multicolumn{1}{c}{ID} & $\lambda_{\rm calc}$ &  $\Vup$ & $\Jup$
& \Vlo & \Jlo & \Elo\ (eV) & \# of lines & $F_{\rm obs}^1$\\
\hline
\hline
3-2 P(4)$^2$ & 1174.923 & 3 & 3 & 2 & 4 & 1.13 & - & -\\  
1-1 P(11) & 1212.425 & 1 & 10 & 1 & 11 & 1.36 & 8 &  39\\  
3-1 P(14) & 1213.356 & 3 & 13 & 1 & 14 & 1.79 & 9 &  42\\  
4-2 R(12) & 1213.677 & 4 & 13 & 2 & 12 & 1.93 & 6 &  19\\  
3-1 R(15) & 1214.465 & 3 & 16 & 1 & 15 & 1.94 & 11 & 140\\  
4-3 P(5) &  1214.781 & 4 & 4  & 3 & 5  & 1.65 & 15 & 96\\  
4-3 R(6) & 1214.995 &  4 & 7  & 3 & 6 & 1.72 & 3 &  8\\  
1-2 R(6) & 1215.726 &  1 & 7  & 2 & 6  & 1.27 & 13 & 162\\  
1-2 P(5) & 1216.070 & 1 & 4 & 2 & 5   & 1.20 & 18 &  350\\  
3-3 R(2) & 1217.031 & 3 & 3 & 3 & 2 & 1.50  & 5 &   17\\  
3-3 P(1) & 1217.038 & 3 & 0 & 3 & 1 & 1.48 & 5 &    33\\  
0-2 R(0) & 1217.205 & 0 & 1 & 2 & 0 & 1.00 & 10 &    395\\  
4-0 P(19) & 1217.410 & 4 & 18 & 0 & 19 & 2.20 & 7 & 36\\  
0-2 R(1) & 1217.643 & 0 & 2 & 2 & 1 & 1.02 & 11 &    360\\  
2-1 P(13) & 1217.904 & 2 & 12 & 1 & 13 & 1.64 & 13 & 179\\  
3-0 P(18) & 1217.982 & 3 & 17 & 0 & 18 & 2.02 & 3 & 10\\  
2-1 R(14) & 1218.521 & 2 & 15 & 1 & 14 & 1.79 & 7 & 22\\  
0-2 R(2) & 1219.089 & 0 & 3 & 2 & 2 & 1.04 & 4&      13\\  
0-2 P(1) & 1219.368 & 0 & 0 & 2 & 1 & 1.02 & 3 &     10\\  
0-5 P(18) & 1548.146 & 0 & 17 & 5 & 18 & 3.78 & 5 & 19\\  
\hline
\multicolumn{9}{l}{$^110^{-15}$ erg cm$^{-2}$ s$^{-1}$}\\
\multicolumn{9}{l}{$^2$Secondary pump - primary pump is 3-3 R(2) at 1217.03}\\
\end{tabular}
\end{table}

\end{center}

\end{document}